\documentclass{article}

\usepackage[final, nonatbib]{neurips_2025}
\usepackage{makecell}
\usepackage[utf8]{inputenc} 
\usepackage[T1]{fontenc}    
\usepackage{hyperref}       
\usepackage{url}            
\usepackage{booktabs}       
\usepackage{amsfonts}       
\usepackage{nicefrac}       
\usepackage{microtype}      
\usepackage{xcolor}         
\usepackage{graphicx}
\usepackage{amsmath}

\usepackage{multirow} 
\usepackage{wrapfig}
\usepackage{subcaption} 
\usepackage{enumitem}
\usepackage[normalem]{ulem}
\useunder{\uline}{\ul}{}
\usepackage{algorithm}
\usepackage{algpseudocode}

\title{SPOT-Trip: Dual-Preference Driven Out-of-Town Trip Recommendation}

\author{
  Yinghui~Liu$^{1}$\thanks{Equal contribution.}, 
  Hao~Miao$^{2}$\footnotemark[1], 
  Guojiang~Shen$^{1}$, 
  Yan~Zhao$^{3}$,  Xiangjie~Kong$^{1}$\thanks{Corresponding author.},
  Ivan~Lee$^{4}$\\
  $^1$Department of Computer Science, Zhejiang University of Technology, Hangzhou, China\\
  $^2$Department of Computing, Hong Kong Polytechnic University, Hong Kong, China\\
  $^3$Shenzhen Institute for Advanced Study, University of Electronic Science and Technology \\of China, Shenzhen, China\\  
  $^4$STEM, University of South Australia, Adelaide,
 Australia\\
  \texttt{\{2112112249, gjshen1975\}@zjut.edu.cn},
  \texttt{hao.miao@polyu.edu.hk},\\
  \texttt{zhaoyan@uestc.edu.cn}, \texttt{xjkong@ieee.org}, \texttt{Ivan.Lee@unisa.edu.au}
}

\begin{document}

\maketitle

\begin{abstract}
Out-of-town trip recommendation aims to generate a sequence of Points of Interest (POIs) for users traveling from their hometowns to previously unvisited regions based on personalized itineraries, e.g., origin, destination, and trip duration. Modeling the complex user preferences--which often exhibit a two-fold nature of static and dynamic interests--is critical for effective recommendations. However, the sparsity of out-of-town check-in data presents significant challenges in capturing such user preferences. Meanwhile, existing methods often conflate the static and dynamic preferences, resulting in suboptimal performance. In this paper, we for the first time systematically study the problem of out-of-town trip recommendation. A novel framework SPOT-Trip is proposed to explicitly learns the dual static-dynamic user preferences. 
Specifically, to handle scarce data, we construct a POI attribute knowledge graph to enrich the semantic modeling of users’ hometown and out-of-town check-ins, enabling the static preference modeling through attribute relation-aware aggregation. 
Then, we employ neural ordinary differential equations (ODEs) to capture the continuous evolution of latent dynamic user preferences and innovatively combine a temporal point process to describe the instantaneous probability of each preference behavior. Further, a static-dynamic fusion module is proposed to merge the learned static and dynamic user preferences. 
Extensive experiments on real data offer insight into the effectiveness of the proposed solutions, showing that SPOT-Trip achieves performance improvement by up to 17.01\%.


\end{abstract}

\section{Introduction} \label{Intro}
With the proliferation of location-based social networks (LBSNs), location-based recommendation has become an important means to help people discover attractive and interesting points of interest (POIs)~\cite{chen2021points, al2024scrollypoi}, including POI recommendation~\cite{yin2023next, kong2024kgnext}, trip recommendation~\cite{gao2023dual, kuo2023bert, zhang2024encoder, shu2024analyzing}, and out-of-town recommendation~\cite{xin2021out, xin2022captor, liu2024kddc}. Traditional POI and trip recommender systems are dedicated to recommending POIs within a specific region. However, they may fail when users travel out of their hometown~\cite{xin2022captor}. 
Consequently, out-of-town recommendation~\cite{liu2024kddc} emerges, which generates POI recommendations for users traveling from their hometowns to regions that they have seldom visited previously.

\begin{wrapfigure}{r}{8.4cm}
\centering
\vspace{-0.4cm}
\includegraphics[width=1\linewidth]{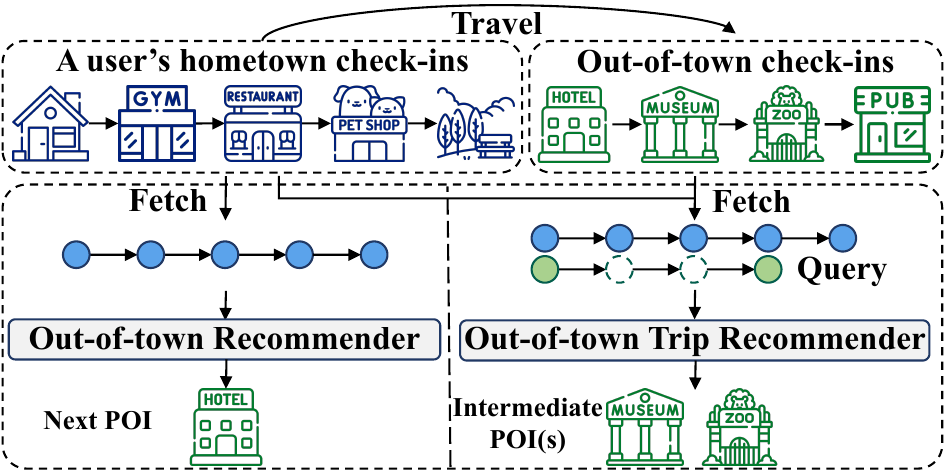}
\caption{Comparison of two out-of-town recommendation tasks. The blue circles indicate the user's hometown historical check-ins, green circles represent the given query POIs, i.e., the origin and destination of a trip, while the hollow dashed circles denote the intermediate POI(s) to be inferred.}
\vspace{-0.3cm}
\label{problem}
\end{wrapfigure}

Existing out-of-town recommendation methods often focus on addressing the problem of interest drift~\cite{wang2017location, liu2024kddc} and geographical gap~\cite{li2020deep, ding2019learning}. However, these methods are mainly designed for the next POI recommendation (see the left part of Fig.~\ref{problem}) while lacking the capabilities to provide a comprehensive trip itinerary for travelers. In real-world scenarios, given an origin and a destination, users may prefer a model that can generate a sequence of intermediate activities (see the right part of Fig.~\ref{problem}) to realize a more engaging journey for convenience. To achieve this, we for the first time study a new problem, i.e., out-of-town trip recommendation, that can provide consecutive intermediate POIs given the origin, destination, and the number of stops. Such trip recommendation models are expected to enable systematic itinerary generation, facilitating efficient decision-making for the users. 


However, \emph{data sparsity} poses a great challenge for out-of-town trip recommendations since users often have few or even no historical check-in records in out-of-town regions. It is hard to obtain a well-performed out-of-town trip recommender model with such scarce data.
In addition, it is challenging to learn the \emph{complex user preferences} to alleviate interest drifts (i.e., out-of-town check-ins are not aligned with hometown check-in preferences) for effective out-of-town trip recommendation~\cite{xin2021out, xin2022captor}. Intuitively, the user preferences can be categorized into two complementary components: \emph{static (or invariant)} and \emph{dynamic preferences}. On the one hand, static preference captures long-term user interests and stable behavioral tendencies, which are often extracted from hometown check-in records. Although static preference shows the stable tendencies of a user, directly applying it to out-of-town trip recommendation may be suboptimal due to cold-start and data scarcity. To enable better generalization, it calls for an alignment method to effectively transfer sufficient knowledge learned from the hometown, i.e., static preferences, to the target regions. On the other hand, dynamic preference reflects the short-term behavioral patterns that are sensitive to contextual information, e.g., time, location, and intent. 

Nonetheless, existing methods~\cite{kong2024kgnext, liu2024kddc} often learn the entangled user preferences. We argue that effective disentanglement of the two preferences can facilitate user intent modeling across diverse scenarios and explicitly mitigate the interest drift, thereby enabling personalized recommendation and improving the robustness and effectiveness. Further, it is challenging to fuse the static and dynamic preferences, which enhances trip recommendation by taking preference consistency and personalization into account, simultaneously, enabling more accurate and context-aware user modeling. To this end, we propose a \underline{S}tatic-dynamic \underline{P}reference aware \underline{O}ut-of-\underline{T}own \underline{Trip} recommendation framework, SPOT-Trip, to explicitly learn such dual user preferences. SPOT-Trip encompasses three major modules: knowledge-enhanced static preference modeling, ODE-based dynamic preference learning, and static-dynamic preference fusion.


The core idea of SPOT-Trip lies in jointly modeling sequence-level static preferences with semantics-enhanced representations and POI-level dynamic preferences by an ODE. 
First, we propose a knowledge-enhanced static preference learning module. In particular, we construct a POI attribute knowledge graph based on user check-ins to incorporate rich semantic relations, e.g., a \textit{hasRating} relation between POIs and attribute entities such as \textit{5-star}. Next, a relation-aware attention aggregation mechanism is designed to generate enriched POI embeddings, which capture the entity semantics with diverse relational contexts, alleviating the data sparsity. Further, we propose a novel static preference alignment mechanism to transfer knowledge of the static preferences learned from hometown check-ins to the out-of-town region. The static preferences are learned by a static aggregator, which aggregates the enriched POI embeddings at the sequence level. 
Second, we develop an ODE-based dynamic preference learning module to model the continuous and irregular evolution based on the dynamic user behaviors during out-of-town trips. 
A temporal point process~\cite{iakovlev2025learning} is further employed to characterize the probability of preference behaviors over time, enabling dynamic preference inference. Notably, we incorporate auxiliary geographical coordinate information to refine the behavior representation. Finally, a static-dynamic preference fusion module is proposed to fuse the learned static and dynamic user preferences for effective out-of-town trip recommendation.

The major contributions of the work are as follows: (1) \emph{New Task}. To the best of our knowledge, this is the first systematic study to learn out-of-town trip recommendation. We propose a framework called SPOT-Trip to explicitly capture the static and dynamic user preferences for effective out-of-town trip recommendation. 
(2) \textit{Novel Techniques}. An innovative static preference learning module is proposed, which leverages sufficient semantic knowledge to extract stable user preferences. We extract the dynamic user preference based on neural ODE in conjunction with a novel temporal point process, which can capture the continuous preference drift.
(3) \textit{Superior Performance}. We report on extensive experiments using real data, offering evidence of the effectiveness of the proposals.

The remainder of this paper is organized as follows. Section~\ref{Preliminary} covers preliminary concepts and formalizes the studied problem. The SPOT-Trip framework and experiments are reported in Sections~\ref{Framework} and~\ref{Experiment}. Section~\ref{conclusion} concludes the paper. For brevity, we survey the related work in Appendix~\ref{RelatedWork}.

\section{Preliminary}
\label{Preliminary}
\textbf{Definition 1: (POI Attribute Knowledge Graph).}
The POI attribute knowledge graph is defined as $\mathcal{G}_k = {(v, r, e)}$, which encodes a semantic relation $r$ between a POI $v$ and an entity $e$. It captures external knowledge by incorporating diverse types of attribute entities and their relationships with POIs, such as (\textit{Balboa Park}, \textit{Located in}, \textit{San Diego}).

\textbf{Definition 2: (Check-in).}
A user check-in is denoted as a tuple $c = (u, t, l, v)$, indicating that user $u$ visited POI $v$ at time $t$ and location $l$, where $l$ includes the geographic coordinates, i.e., latitude and longitude.

\textbf{Definition 3: (Out-of-town Travel Behavior).}
Given a user $u$, the out-of-town travel behavior is denoted as $\xi = \bigl(u, \vec{c}_h, \vec{c}_o, a_h, a_o\bigr)$, indicating that $u$ departs from his/her hometown $a_h$ to visit an out-of-town region $a_o$ with check-in records in both regions, i.e., $\vec{c}_h$ (hometown) and $\vec{c}_o$ (out-of-town), respectively. We use $ M = |\vec{c}_h| $ and $ N = |\vec{c}_o| $ to represent the number of check-ins in the hometown and out-of-town regions, respectively, where $ M > N$.


\textbf{Problem Statement: (Out-of-town Trip Recommendation).}
Given a set of users $\mathcal{U} = \{u_i\}_{i=1}^{|\mathcal{U}|}$, POIs $\mathcal{V} = \{v_i\}_{i=1}^{|\mathcal{V}|}$, regions $\mathcal{A} = \{a_i\}_{i=1}^{|\mathcal{A}|}$, and out-of-town trip records $\mathcal{O} = \{\xi_i\}_{i=1}^{|\mathcal{O}|}$, our objective is to learn a recommender function $\mathcal{F}$ based on historical records $\mathcal{O}$ and the POI attribute knowledge graph $\mathcal{G}_k$. For a user $u^* \notin \mathcal{U}$ at $a_h^*$ with hometown check-ins $\vec{c}_{h}$ and out-of-town origin-destination trip query $Q_o^* = \{ v_{s}^o, v_{e}^o, N \}$ in region $a_o^*$, where $v_{s}^o$, $v_{e}^o$, and $N$ denote the start, end points and the number of stops, the learned $\mathcal{F}$ generates a sequence of POIs $\tau = \{ v_1^o, v_2^o, \dots, v_N^o \}$, where $v_1^o = v_s^o$ and $v_N^o = v_e^o$ for $u^*$. The recommended POIs in $\tau$ are in $a_o^*$. 

\section{Methodology}
\label{Framework}
\begin{figure*}[tb]
  \centering
  \includegraphics[width=\textwidth]{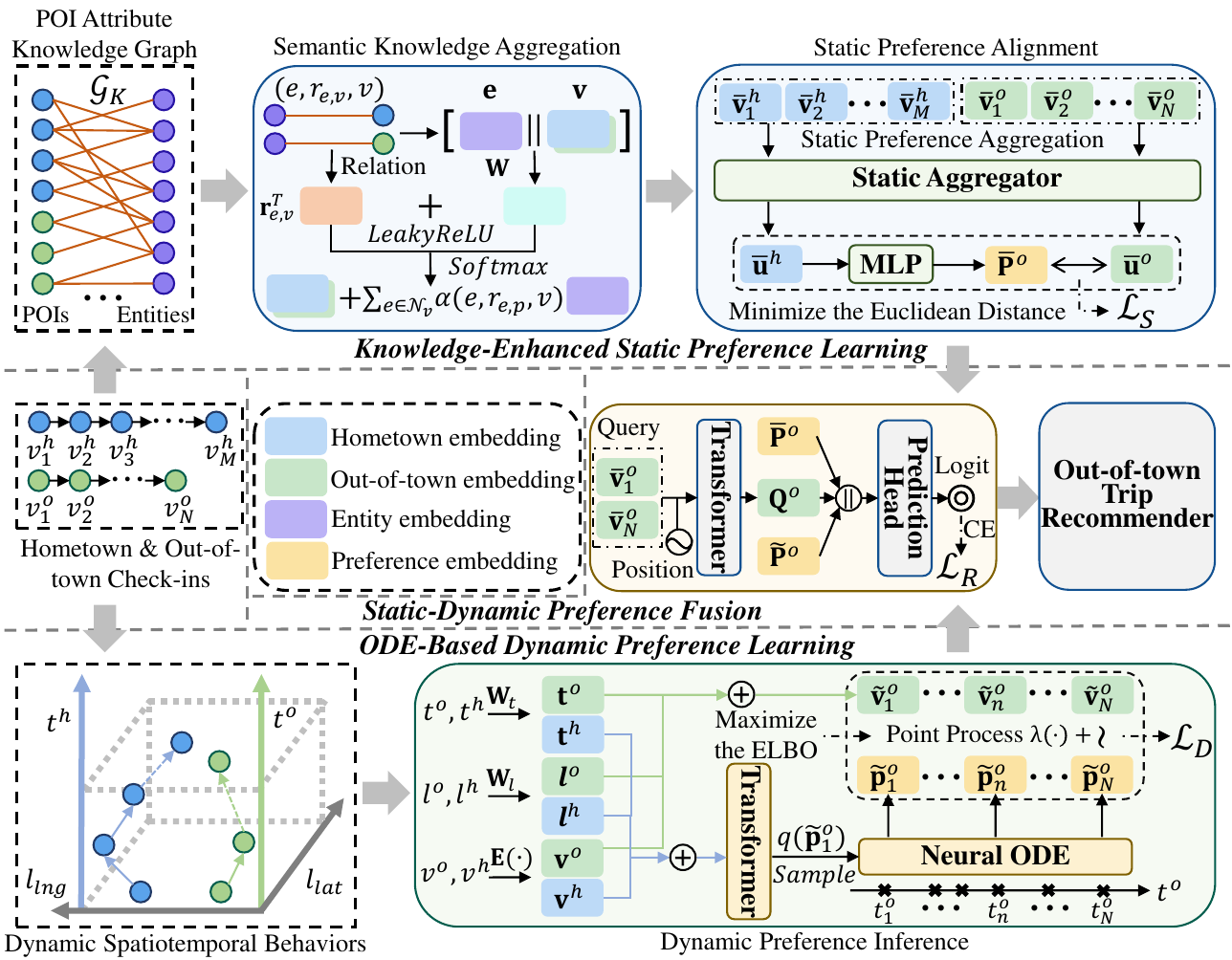}
  \caption{The framework of SPOT-Trip. CE denotes the cross-entropy loss.}
  \label{framework}
  \vspace{-0.4cm}
\end{figure*}

We proceed to detail the dual preference-aware out-of-town trip recommendation framework, SPOT-Trip. We first give an overview of the framework and then provide specifics. As illustrated in Fig.~\ref{framework}, SPOT-Trip encompasses three major modules: Knowledge-Enhanced Static Preference Learning (\textit{KSPL}), ODE-Based Dynamic Preference Learning (\textit{ODPL}), and Static-Dynamic Preference Fusion.

\textit{KSPL} aims to enhance the static semantic awareness of user preference by means of knowledge graph construction, which consists of two components. First, semantic knowledge aggregation is proposed to improve POI embeddings with a relational attention mechanism based on the constructed POI attribute knowledge graph. Second, we design an innovative static preference alignment component that facilitates the transfer of static hometown preferences to the out-of-town region. \textit{ODPL} is dedicated to capturing the evolving nature of user preferences with a novel ODE-based continuous dynamics modeling. 
In this module, we first aggregate the hometown check-in sequence with spatiotemporal information and feed the resulting sequence into a Transformer to generate initial latent representations, which are then input into a neural ODE to capture dynamics. Subsequently, the check-in behaviors are further interpreted by a temporal point process to characterize the instantaneous probability of each behavior. Finally, we propose a static-dynamic preference fusion module to merge the learned static and dynamic user preferences.

\subsection{Knowledge-Enhanced Static Preference Learning} \label{sec:3.1}
Traditional out-of-town recommendation methods have attempted to model static preference with transition relationships~\cite{xin2021out} and geographic relationships~\cite{xin2022captor}. However, they remain hampered by data sparsity and fail to align user preferences across different regions. Knowledge graphs (KGs)~\cite{cao2024knowledge, zhang2024survey} enrich POI embeddings using additional semantic information, thereby strengthening user preference representations. Yet, directly integrating semantic and spatial information may lead to conflicting interactions~\cite{liu2024kddc}. Accordingly, we leverage a KG approach solely to model the static user preferences from a semantic perspective. Spatial information will be incorporated in the dynamic learning described in Sec.~\ref{sec:3.2}.

\textbf{Semantic Knowledge Aggregation.} 
In the POI attribute knowledge graph, various relationships (e.g., POI's category and corresponding region) encode distinct semantic information. Inspired by the graph attention mechanisms in~\cite{yang2023knowledge, liu2024kddc}, we first introduce a relation-aware knowledge embedding layer to reflect the heterogeneity of relations over knowledge graph connections and then gain entity- and relation-specific representations through a parameterized attention mechanism. Towards that, we construct our relation-aware message aggregation mechanism between the POI and its connected entities in $\mathcal{G}_K$, for generating knowledge-aware POI embeddings via a heterogeneous attentive aggregator illustrated as follows:
\begin{equation}
\label{eq:1}
\bar{\mathbf{v}} = \mathbf{v} + \sum_{e \in \mathcal{N}_v}\alpha (e,r_{e,v},v)\mathbf{e}, 
\alpha (e,r_{e,v},v) = \frac{\mathrm{exp}(\phi(\mathbf{r}_{e,v}^{T} \mathbf{W} [\mathbf{e}\parallel \mathbf{v}]))}{ {\textstyle \sum_{e\in \mathcal{N} _{v}} } \mathrm{exp}(\phi(\mathbf{r}_{e,v}^{T} \mathbf{W} [\mathbf{e}\parallel \mathbf{v}]))},
\end{equation}
where $\mathcal{N}_v$ is the neighboring entities of POI $v$ on various relations $r_{e,v}$ in $\mathcal{G}_K$, $\mathbf{v} \in \mathbb{R}^d$ and $\mathbf{e} \in \mathbb{R}^d$ represent the embedding of POI and entity, respectively. The estimated entity- and relation-dependent attentive relevance during the knowledge aggregation process is denoted as $\alpha (e,r_{e,v},v)$, which encodes the distinct semantics of relationships between POI $v$ and entity $e$. $\parallel$ means the concatenation of two embeddings. $\mathbf{W} \in \mathbb{R}^{d \times 2d}$ denotes the weight matrix that is tailored to the input POI and entity representations. $\phi$ is the activation function \textit{LeakyReLU} for non-linear transformation. Additionally, to further improve the multi-relational semantic representation space for entity–item dependencies, an alternative training is employed between the relation-aware knowledge aggregator and TransE~\cite{bordes2013translating} (More details about the alternative training and its loss can be found in the Appendix~\ref{Alternative Training}).

\textbf{Static Preference Alignment.} Inspired by prior out-of-town recommendation approaches that derive user preferences through behavioral aggregation, we utilize a static aggregator $\mathrm{AGG}_S$(·) (i.e., average pooling) to consolidate the enriched POI embeddings, producing enhanced user hometown and out-of-town static preference representations:
$
\bar{\mathbf{u}}^h = \mathrm{AGG}_S([\bar{\mathbf{v}}_{m}^{h}]_{m=1}^{M} ), \bar{\mathbf{u}}^o = \mathrm{AGG}_S([\bar{\mathbf{v}}_{n}^{o}]_{n=1}^{N} ).
$
Subsequently, we infer user out-of-town preference by a MLP based on their hometown preferences:
\begin{equation} \label{eq:2}
\bar{\mathbf{P}}^o=\phi (\mathbf{W}_{S}\bar{\mathbf{u}}^h + \mathbf{b}_{S} ),   
\end{equation}
where $\mathbf{W}_{S} \in \mathbb{R}^{d \times d}$ and $\mathbf{b}_{S} \in \mathbb{R}^{d}$ are trainable parameters. $\phi$ denotes the activation function \textit{SiLU}. An Euclidean distance loss $\mathcal{L}_{S}$ is adopted to bridge the inferred preference and the actual out-of-town preference for static learning as follows:
\begin{equation}
\label{eq:3}
\mathcal{L}_{S} = \sum_{u \in \mathcal{U}}  \left \| \bar{\mathbf{P}}^o-\bar{\mathbf{u}}^o \right \| _{2}^{2}
\end{equation}
\subsection{ODE-Based Dynamic Preference Learning}
\label{sec:3.2}
Previous trip recommenders~\cite{kuo2023bert, shu2024analyzing} incorporate position embeddings and periodic hour encodings to model the dynamic nature of user preferences. However, due to the irregular sampling~\cite{qin2024learning} of check-in data, these approaches fail to capture the dynamic evolution of user preferences over actual time. Given the success of neural ODE~\cite{chen2018neural} in other research fields~\cite{mercatali2024graph, iakovlev2025learning}, we develop a neural ODE to model the continuous dynamic drift of user out-of-town preferences in latent space. To quantify the behavior event probability at each moment, we also formulate the preference inference as a temporal point process~\cite{ludke2023add}, where the instantaneous probability is described by the intensity function $\lambda(\cdot)$ of a non-homogeneous Poisson process (NHPP) on the latent state. Overall, the \textit{Dynamic Preference Inference} process can be defined as:
\begin{equation}
\label{eq:4}
\tilde{\mathbf{p}}^o_{t_1} \sim p\bigl(\tilde{\mathbf{p}}^o_{t_1}\bigr),
\frac{d\tilde{\mathbf{p}}^o_t}{dt} = f\bigl(\tilde{\mathbf{p}}^o_{t_1}\bigr),
\end{equation}
\begin{equation}
\label{eq:5}
t_n \sim \mathrm{NHPP}(\lambda(\tilde{\mathbf{p}}^o_{t})),
n=1,...,N,
\end{equation}
\begin{equation}
\label{eq:6}
\tilde{\mathbf{v}}^o_{t_n} \sim p\bigl(\tilde{\mathbf{v}}^o_{t_n}|\tilde{\mathbf{p}}^o_{t_n}\bigr),
n=1,...,N,
\end{equation}
where $\tilde{\mathbf{p}}^o_{t}$ represents the latent preference state at time $t$. $f(\cdot)$ and $\lambda(\cdot)$ are MLPs for parameterizing ODE dynamics and the intensity function. For brevity, we denote  $\tilde{\mathbf{p}}^o_{t_n}$ by $\tilde{\mathbf{p}}^o_n$ 
over the remainder of this paper, and the corresponding joint distribution is:
\begin{align} \label{eq:7}
&p(\tilde{\mathbf{p}}^o_1, \{t_{n},\tilde{\mathbf{v}}^o_n\}^N_{n=1})=p(\tilde{\mathbf{p}}^o_1)p(\{t_{n}\}^N_{n=1}|\tilde{\mathbf{p}}^o_1)\prod_{n=1}^{N}p(\tilde{\mathbf{v}}^o_n|\tilde{\mathbf{p}}^o_n,t_{n}),p(\tilde{\mathbf{p}}^o_1) = \mathcal{N}(\mathbf{0},I), \nonumber \\ 
& p(\{t_{n}\}^N_{n=1}|\tilde{\mathbf{p}}^o_1) = \prod_{n=1}^{N}\lambda(\tilde{\mathbf{p}}^o_n)\mathrm{exp}(-\int\lambda(\tilde{\mathbf{p}}^o)dt), p(\tilde{\mathbf{v}}^o_n|\tilde{\mathbf{p}}^o_n,t_{n})= \mathcal{N}(\tilde{\mathbf{p}}^o_n, \sigma^2_{\tilde{v}^o} I), 
\end{align}
where $\mathcal{N}$ is the normal distribution, $\mathbf{0}$ is a zero vector, $I$ is the identity matrix and $\sigma_{\tilde{v}^o}$ denotes a hyper-parameter that sets the model's tolerance for discrepancies between observations and predictions. In the following, the detailed descriptions and inferences of each component are provided.

\textbf{Latent Dynamics of Preference Drift (Eq.~\ref{eq:4}).}
Before modeling the latent ODE dynamics, the user's dynamic spatiotemporal behaviors including timestamps and spatial coordinates are first mapped into the latent space via linear transformations and a new embedding layer $\mathbf{E}(\cdot)$: 
$\tilde{\mathbf{v}}^{h,o}=\mathbf{W}_tt^{h,o} + \mathbf{W}_ll^{h,o} + \mathbf{E}(v^{h,o})$, 
where $\mathbf{W}_{t} \in \mathbb{R}^{1 \times d}$ and $\mathbf{W}_{l} \in \mathbb{R}^{2 \times d}$ are trainable parameters. Note that the POIs' spatial features are treated solely as quantitative values because geographic information is unavailable during the recommendation. To better infer the posterior of the latent initial state of preference $\tilde{\mathbf{p}}^o_{1}$ and model parameters, the variational inference~\cite{agrawal2021amortized} is used for latent modeling. We define an approximate posterior $q(\tilde{\mathbf{p}}^o_1;\psi)$ with variational parameters $\psi$ to approximate the true posterior $p(\tilde{\mathbf{p}}^o_{1}|\{t_n^o,\tilde{\mathbf{v}}_n^o\}_{n=1}^N)$, with the objective of minimizing the Kullback-Leibler divergence 
\begin{equation}
\label{eq:8}
\mathrm{KL}[q(\tilde{\mathbf{p}}^o_1;\psi)||p(\tilde{\mathbf{p}}^o_{1}|\{t_n^o,\tilde{\mathbf{v}}_n^o\}_{n=1}^N)]  
\end{equation}
over the variational parameters to derive an estimate of the approximation of the posterior and model parameters.
To avoid optimizing the local variational parameters $\psi$ for each behavior trajectory in the dataset, we use amortization~\cite{agrawal2021amortized} and define $\psi$ as a reparameterization sampling from the output of a stacked Transformer encoder layer~\cite{vaswani2017attention} $\mathrm{Trans}_D$. To be specific, we first introduce an aggregation token, $\mathbf{AGG}$, with a learnable representation to indicate $\mathrm{Trans}_D$ to aggregate user hometown behaviors. This gives us the encoder input sequence: $\{\{\tilde{\mathbf{v}}^{h}_m\}_{m=1}^M,\mathbf{AGG}\}$, and then we acquire the output aggregation token $\tilde{\mathbf{AGG}}$ at the last layer. Finally, we obtain $\psi$ from $\tilde{\mathbf{AGG}}$ with the reparameterization sampling dependent on an approximate posterior (see Appendix~\ref{Posterior} for specification of the sampling).

\textbf{Temporal Point Process (Eq.~\ref{eq:5}).}
After receiving the inferred preference $\tilde{\mathbf{p}}^o_{n}$ from the ODE solver in Eq.~\ref{eq:4}, we employ the parameterized intensity function $\lambda$(·) to map it to the intensity of NHPP. In order to ensure that the intensity function value is nonnegative and enhance its numerical stability, we further exponentiate the output of $\lambda(\tilde{\mathbf{p}}^o_{n})$ and then add a small constant.

\textbf{Behavior Distribution Reconstruction (Eq.~\ref{eq:6}).} In practice, rather than minimizing the KL divergence in Eq.~\ref{eq:8} directly, we instead maximize the corresponding evidence lower bound (ELBO) for the dynamic behavior distribution reconstruction, and the total dynamic learning loss is defined as:
\begin{align}\label{eq:9}
\mathcal{L}_D=-\sum_{u \in \mathcal{U}}(&  \underbrace{\sum_{n=1}^{N_u}\mathbb{E}_{q(\tilde{\mathbf{p}}^o_{1};\psi)}[\ln p(\tilde{\mathbf{v}}^{o}_n|t_n^o,\tilde{\mathbf{p}}^o_{1})]}_{(i)\text{ Expected restructured behavior log-lik}} 
+ \underbrace{\mathbb{E}_{q(\tilde{\mathbf{p}}^o_{1};\psi)}\sum_{n=1}^{N_u}\ln \lambda(\tilde{\mathbf{p}}^o_{n})-\int\lambda(\tilde{\mathbf{p}}^o)dt}_{(ii)\text{ Expected temporal point process log-lik}}  \nonumber \\
&- \underbrace{\mathrm{KL}[q(\tilde{\mathbf{p}}^o_1;\psi)||p(\tilde{\mathbf{p}}^o_{1})]}_{(iii)\text{ KL between prior and posterior}}),
\end{align}
where the term (iii) can be computed analytically, computation of terms (i) and (ii) involves approximations: term (i) is estimated via Monte Carlo integration using samples from the variational posterior, whereas term (ii) involves solving an ODE with a numerical solver and applying the trapezoidal rule to approximate the expected intensity integral. The ELBO is maximized w.r.t. the model parameters and its detailed derivation is contained in Appendix~\ref{ELBO}. 
\subsection{Static-Dynamic Preference Fusion} \label{fusion}
To generate the logit $z_{u,n,i}$ for user $u$ at position $n$ over each POI $i \in a_o$, we first construct a unified representation by combining the user's knowledge-enhanced query representation $\mathbf{Q}^o  \in \mathbb{R}^{2d \times d}$, the static preference $\bar{\mathbf{P}}^o$ inferred in Sec.~\ref{sec:3.1}, and the dynamic preference sequence $\tilde{\mathbf{P}}^o = \{\tilde{\mathbf{p}}^o_n\}_{n=1}^N$ inferred in Sec.~\ref{sec:3.2}. This joint representation is then fed into a nonlinear prediction head to compute the logits for all POIs at each position $n$ in the out-of-town region $a_o$. 
\begin{equation}\label{eq:10}
z_{u,n,i} = \phi(\mathbf{W}_R[\mathbf{Q}^o||\bar{\mathbf{P}}^o||\tilde{\mathbf{P}}^o]+\mathbf{b}_R),
\end{equation}
where $\mathbf{W}_{R} \in \mathbb{R}^{4d \times d}$, $\mathbf{b}_{R} \in \mathbb{R}^{d}$ are trainable parameters and $\sigma$ is the activation function \textit{LeakyReLU}.
Appendix~\ref{ELBO} contains more details about the acquisition of query representation $\mathbf{Q}^o$. Consistent with previous works~\cite{gao2023dual, shu2024analyzing}, the main recommendation loss is formulated as a cross-entropy loss:
\begin{equation}\label{eq:11}
\mathcal{L}_R= -\frac{1}{\sum_{u\in \mathcal{U}} N_u} \sum_{u=1}^{\mathcal{U}}\sum_{n=1}^{N_u} \log \frac{\exp\bigl(z_{u,n,v^o_{u,n}}\bigr)}{\sum_{i\in a_o} \exp\bigl(z_{u,n,i}\bigr)}.
\end{equation}
\subsection{Optimization and Recommendation} \label{Recommendation}
\textbf{Optimization.} With Eqs.~\ref{eq:3},~\ref{eq:9} and~\ref{eq:11}, we can jointly optimize the composite 
training optimization loss function in an end-to-end fashion as below:
\begin{equation}
\label{eq:12}
\begin{aligned}
\mathcal{L} = \beta_{1} \mathcal{L}_{S} + \beta_{2} \mathcal{L}_{D} + \beta_{3} \mathcal{L}_{R},
\end{aligned}
\end{equation}
where $\beta_{1}$, $\beta_{2}$ and $\beta_{3}$ are the hyper-parameters to balance the effects of the three losses in SPOT-Trip.

\textbf{Recommendation.} When carrying out the out-of-town trip recommendation, our goal is to recommend a sequence of out-of-town POIs to a new user $u^*$ traveling to region $a$ with a clear query. In particular, given $u^*$'s hometown check-in records and query $Q_o^*=\{v_s^o,v_e^o,N\}$, we first derive her/his inferred static preference $\bar{\mathbf{P}}^o$ and dynamic preference $\tilde{\mathbf{P}}^o$ following Eqs.~\ref{eq:2} and~\ref{eq:4}. Then we utilize Eq.~\ref{eq:10} to generate each POI logit $z_{u^*,n,i}$ at the trip position $n$ in region $a$ and employ a stochastic sampling method (i.e., Top-p~\cite{Holtzman2020The}) to recommend the intermediate POI(s) for trip completion. Note that actual timestamps are employed in Sec.~\ref{sec:3.2} during training to fully capture the temporal dynamics; however, such precise time information is not available during the recommendation phase. Consequently, we adopt the normalization of each query position $n$ as a surrogate time grid, thereby approximating temporal progression in a computationally efficient manner.
\section{Experiment}
\label{Experiment}
\subsection{Experimental Setups} \label{setups}
\textbf{Dateset.} The experiments are carried out on two widely-used travel behavior datasets: Foursquare and Yelp.
We follow existing studies~\cite{xin2022captor, liu2024kddc} to identify users with check-in activities in their hometown and other regions. The check-ins are reorganized to form the corresponding out-of-town travel records $\bigl(u, \vec{c}_h, \vec{c}_o, a_h, a_o\bigr)$. To improve the rationality, we filter out users who have fewer than three check-ins in out-of-town regions, or whose travel durations are shorter than 1 hour or longer than 30 days. The knowledge graphs are extracted from the supplementary descriptive information in two datasets. More details regarding datasets can be seen in Appendix~\ref{Dataset details}.

\textbf{Baseline.}
We compare SPOT-Trip with 9 baselines, including 6 established trip recommendation methods without hometown information: Popularity~\cite{chen2016learning}, POIRank~\cite{chen2016learning}, GraphTrip~\cite{gao2023dual}, MatTrip~\cite{zhang2024encoder}, AR-Trip~\cite{shu2024analyzing}, and Base; and 3 methods with hometown information: Base + KDDC~\cite{liu2024kddc}, Base + CNN-ODE~\cite{iakovlev2025learning}, and Base + PPROC~\cite{iakovlev2025learning}, where Base denotes SPOT-Trip without \textit{KSPL} and \textit{ODPL}. More baseline details are provided in Appendix~\ref{Baseline details}.


\textbf{Evaluation Metrics.}
Following existing studies~\cite{gao2023dual, kuo2023bert, shu2024analyzing}, $F_1$ and $PairsF_1$, are adopted as the evaluation metrics, where higher values indicate better performance. As the out-of-town trip recommendation focuses on intermediate POIs, we calculate $F_1$ and $PairsF_1$ without the origin and destination. In addition, to enable more comprehensive comparison, we also report the experiment results with origin and destination, denoted as $Full$-$F_1$ and $Full$-$PairsF_1$. Please see the corresponding formulas in  Appendix~\ref{metrics}.


\textbf{Implementation Details.}
We implement our model using the Pytorch framework on NVIDIA GeForce RTX 4090 GPU. We perform statistical testing, i.e., three times, with different parameters to enable fairer comparison, where averaged results are reported in our study. The learning rate is set to $0.001$ with Adam optimizer. Additionally, the batch size and training epochs are set to $32$ and $1000$, respectively.
We provide more details about implementation in Appendix~\ref{more implementation details}. Further, the parameters of the baseline methods are set based on their original papers and associated code.
\begin{table}[]
\centering
\Huge
\caption{The overall comparison between SPOT-Trip and baselines, where the best performance is marked in bold while the second-best results are underlined. \textbf{*} denotes improvements that are statistically significant, where we use two-sided t-test with $p$\text{-}value < 0.05~\cite{liu2024llm}.} 
\label{Comparison}
\resizebox{\columnwidth}{!}{%
\begin{tabular}{@{}c|cccc|cccc@{}}
\toprule
\multirow{2}{*}{Method} & \multicolumn{4}{c|}{Foursquare} & \multicolumn{4}{c}{Yelp} \\
\cmidrule(l){2-9}
& $F_1(\uparrow)$ & $PairsF_1(\uparrow)$ & $Full$-$F_1(\uparrow)$ & $Full$-$PairsF_1(\uparrow)$ & $F_1(\uparrow)$ & $PairsF_1(\uparrow)$ & $Full$-$F_1(\uparrow)$ & $Full$-$PairsF_1(\uparrow)$ \\
\midrule
Popularity~\cite{chen2016learning}  & 0.0261 & 0.0013 & 0.4423    & 0.1565    & 0.0257 & 0.0056 & 0.5065    & 0.2058    \\
POIRank~\cite{chen2016learning}     & 0.0253 & 0.0019 & 0.4416    & 0.1582    & 0.0264 & 0.0079 & 0.5068    & 0.2093    \\
GraphTrip~\cite{gao2023dual}        & 0.0295 & 0.0048 & 0.4498    & 0.1620    & 0.0289 & 0.0126 & 0.5107    & 0.2184    \\
MatTrip~\cite{zhang2024encoder}     & 0.0311 & 0.0037 & 0.4530    & 0.1656    & 0.0301 & 0.0119 & 0.5117    & 0.2191    \\
AR-Trip~\cite{shu2024analyzing}     & 0.0304 & 0.0045 & 0.4512    & 0.1673    & 0.0307 & 0.0153 & 0.5115    & 0.2204    \\ \midrule
Base                                & 0.0339 & 0.0069 & 0.4571    & 0.1698    & 0.0315 & 0.0149 & 0.5097    & 0.2215    \\
\midrule
Base + KDDC~\cite{liu2024kddc}      & \underline{0.0375} & 0.0079 & \underline{0.4606}    & 0.1822    & \underline{0.0341} & 0.0156 & \underline{0.5126}    & \underline{0.2256}    \\ \midrule
Base + CNN-ODE~\cite{iakovlev2025learning} & 0.0367 & \underline{0.0094} & 0.4578    & \underline{0.1843}   & 0.0326 & \underline{0.0168} & 0.5124    & 0.2237    \\
Base + PPROC~\cite{iakovlev2025learning}   & 0.0330 & 0.0071 & 0.4550    & 0.1687    & 0.0334 & 0.0159 & 0.5110    & 0.2218    \\ \midrule
SPOT-Trip                                & $\textbf{0.0400}^*$ & $\textbf{0.0109}^*$ & $\textbf{0.4723}^*$    & $\textbf{0.1960}^*$   & $\textbf{0.0399}^*$ & $\textbf{0.0190}^*$ & $\textbf{0.5261}^*$    & $\textbf{0.2347}^*$    \\ 
Improvement                                & +6.67\% & +15.96\% & +2.54\%    & +6.34\%    & +17.01\% & +13.90\% & +2.63\%    & +4.03\%    \\
\bottomrule
\end{tabular}%
}
\end{table}
\subsection{Overall Performance}
We compare SPOT-Trip with 9 baselines, where Table~\ref{Comparison} reports the $F_1$, $PairsF_1$, $Full\text{-}F_1$, and $Full\text{-}PairsF_1$ values. Generally, SPOT-Trip achieves the best results on both datasets across all evaluation metrics, demonstrating its effectiveness. SPOT-Trip performs better than the best among the baselines by up to 17.01\% and 15.96\% in terms of $F_1$ and $PairsF_1$, respectively. We also observe that the performance improvements obtained by SPOT-Trip on the Yelp dataset exceed those on Foursquare in terms of intermediate POIs recommendation. This is because the sufficient semantic information of Yelp introduces more external knowledge for model training, enhancing model performance. 
Additionally, we observe that out-of-town trip recommendation-based methods that use hometown information generally yield higher $F_1$ and $PairsF_1$ scores than those trip recommendation-based methods without hometown information, suggesting the significance of the static and dynamic user preference modeling with hometown context. 

Further, compared to suboptimal baselines equipped with only a single learning module, i.e., Base + KDDC, Base + CNN-ODE and Base + PPROC, SPOT-Trip achieves notable improvements. Specifically, on Foursquare, it improves the $F_1$ and $PairsF_1$ by 6.67\% and 15.96\%, respectively, while the corresponding gains are 17.01\% and 13.90\% on Yelp. Moreover, under full‑trip evaluations $Full$-$F_1$ and $Full$-$PairsF_1$, which inherently include fixed origin and destination and therefore dilute relative gains, SPOT-Trip still consistently outperforms the baselines. These observations demonstrate the superiority of SPOT-Trip due to that it can learn static and dynamic user preferences comprehensively. It is worth noting that Base + PPROC exhibits inferior performance compared to Base + CNN-ODE, potentially because of the complexity involved in predicting spatiotemporal event points, which largely depends on the design of the underlying predictive model.
\subsection{Ablation Study} \label{ablation}
\begin{wrapfigure}{r}{6cm}
\vspace{-0.3cm}
  \centering
  \includegraphics[width=\linewidth]{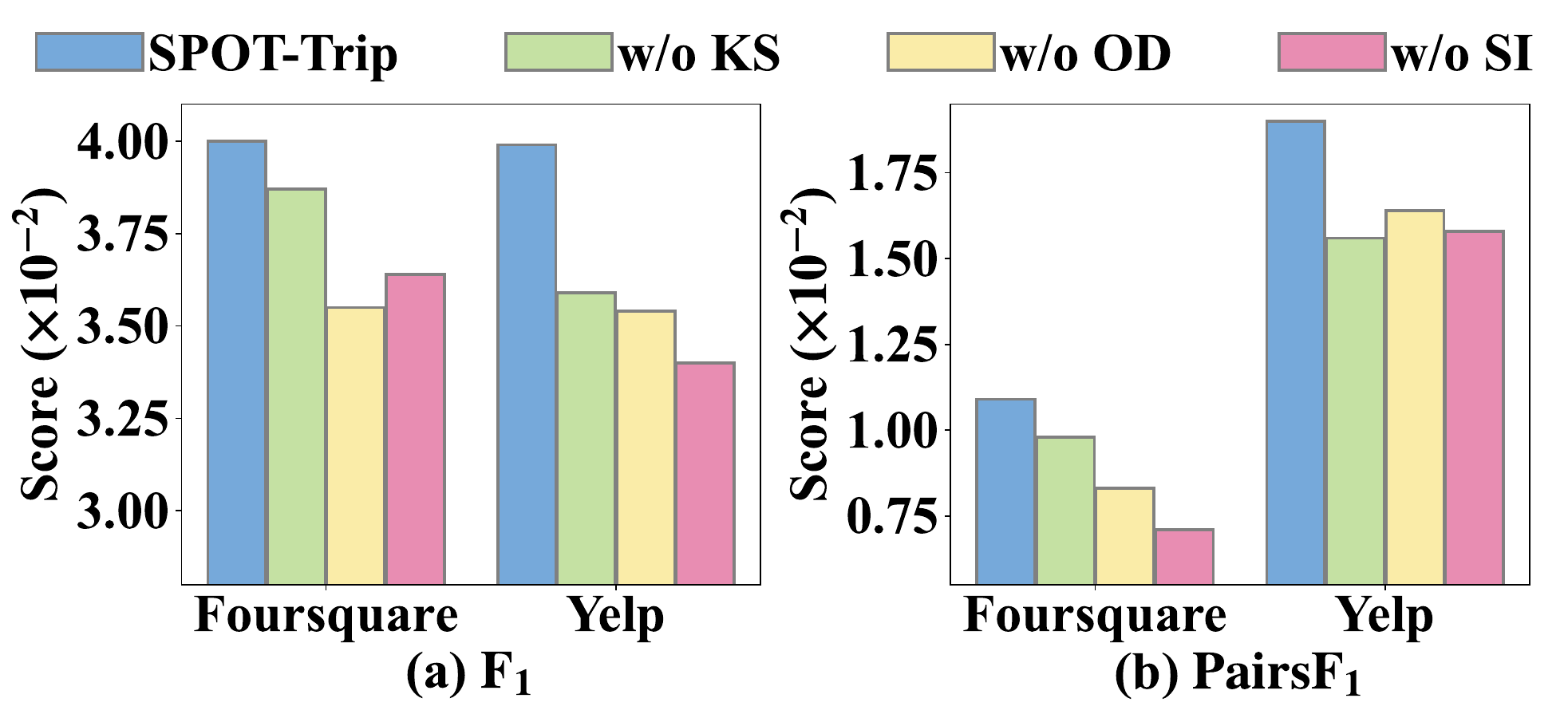}
  \caption{Performance of SPOT-Trip and its variants on two datasets.}
\vspace{-0.3cm}
  \label{Ablation 1}
\end{wrapfigure}
To gain insight into the contributions of the different components of SPOT-Trip, we evaluate three variants: (1) \textbf{w/o KS}. SPOT-Trip without the \textit{KSPL} module; (2) \textbf{w/o OD}. SPOT-Trip without the \textit{ODPL} module; 
(3) \textbf{w/o SI}. SPOT-Trip without the spatial information in \textit{ODPL}. Figure~\ref{Ablation 1} shows the results on two datasets, SPOT-Trip outperforms its counterparts with the \textit{KSPL} module, \textit{ODPL} module, and spatial information. This shows that these three components are useful for effective out-of-town trip recommendations. In particular, the \textbf{w/o KS} variant shows a 10.02\% drop in $F_1$ and a 21.79\% drop in $PairsF_1$ on Yelp, validating the effectiveness of our framework in modeling users’ static preferences. Furthermore, the \textbf{w/o SI} variant suffers significant performance degradation, especially in terms of $PairsF_1$ on Foursquare and $F_1$ on Yelp, demonstrating the necessity of incorporating spatial information into the static preference learning process.
More results and discussions are provided in Appendix~\ref{more ablation}.
\begin{figure}[h]
	\centering
	\begin{subfigure}{0.245\linewidth}
		\centering
		\includegraphics[width=1\linewidth]{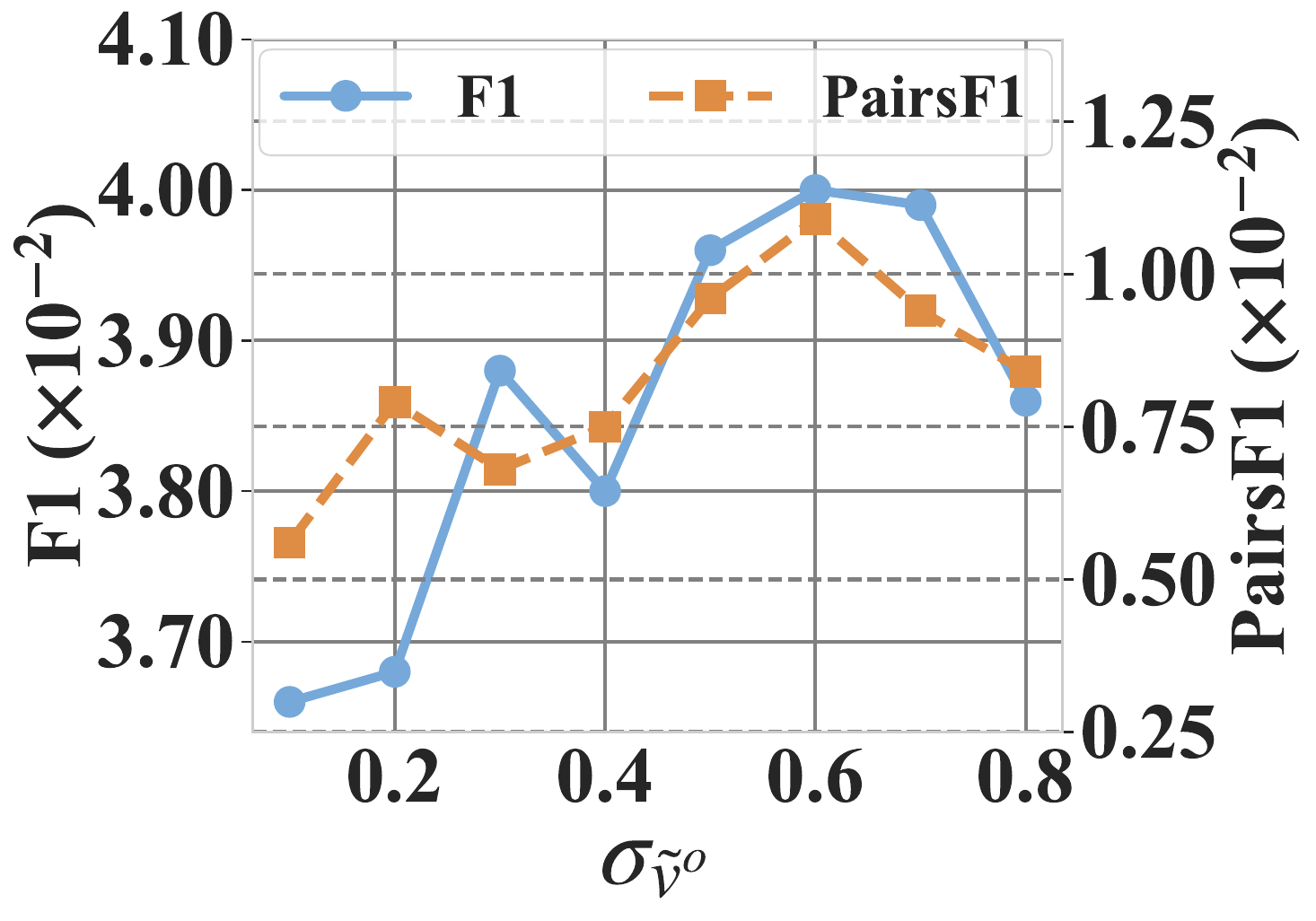}
	\end{subfigure}
	\centering
	\begin{subfigure}{0.245\linewidth}
		\centering
		\includegraphics[width=1\linewidth]{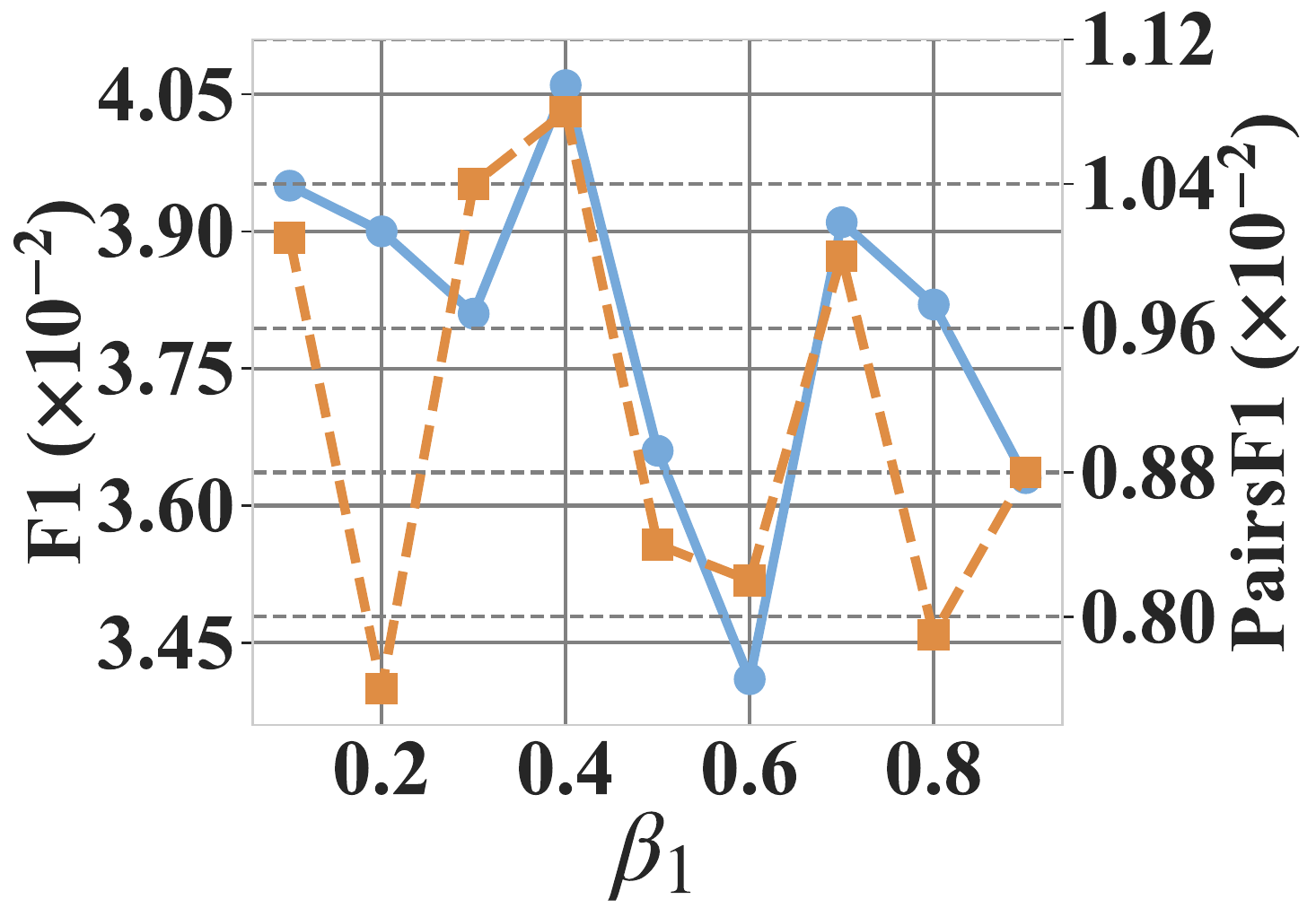}
	\end{subfigure}
	\begin{subfigure}{0.245\linewidth}
		\centering
		\includegraphics[width=1\linewidth]{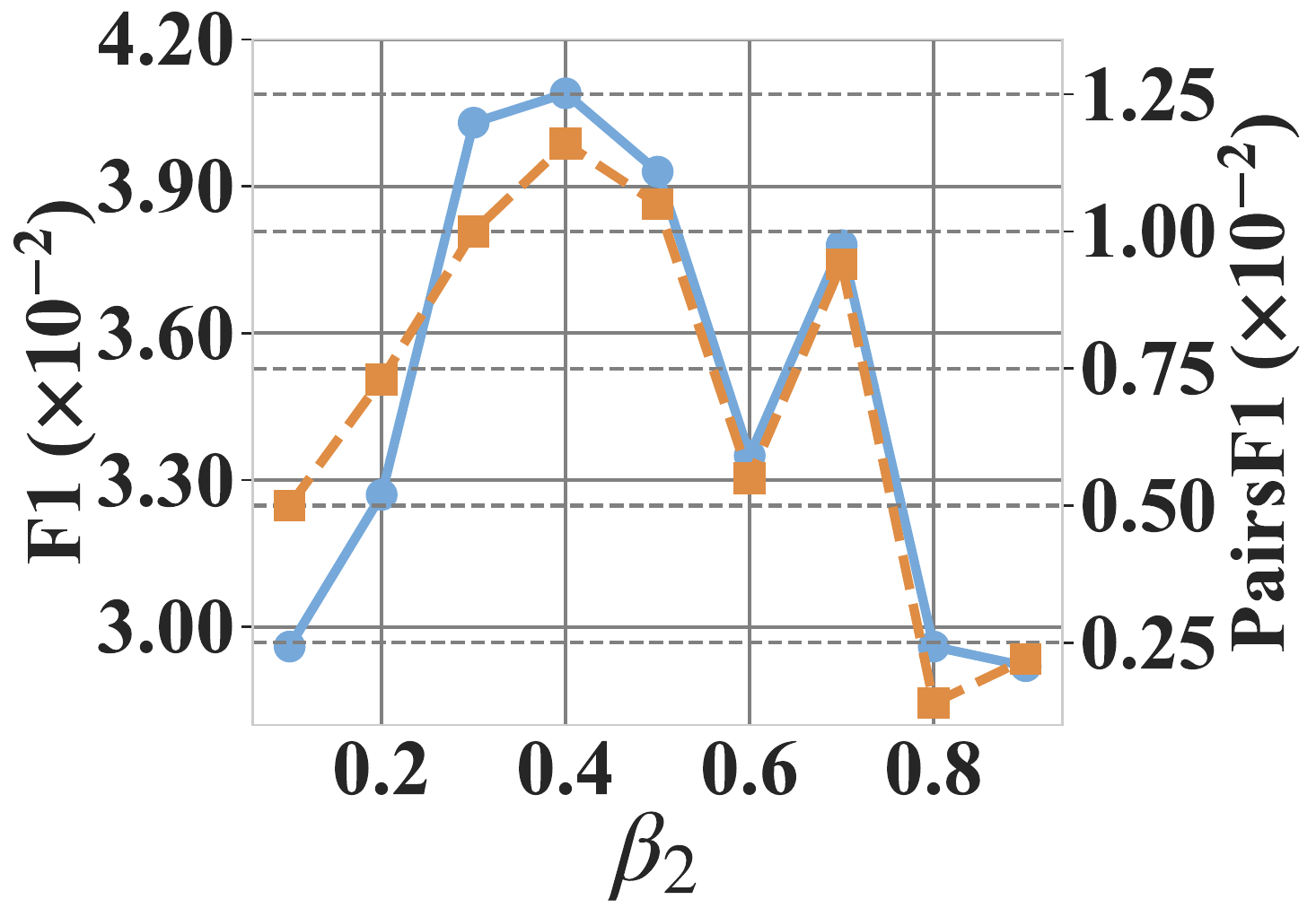}
	\end{subfigure}
	\begin{subfigure}{0.245\linewidth}
		\centering
		\includegraphics[width=1\linewidth]{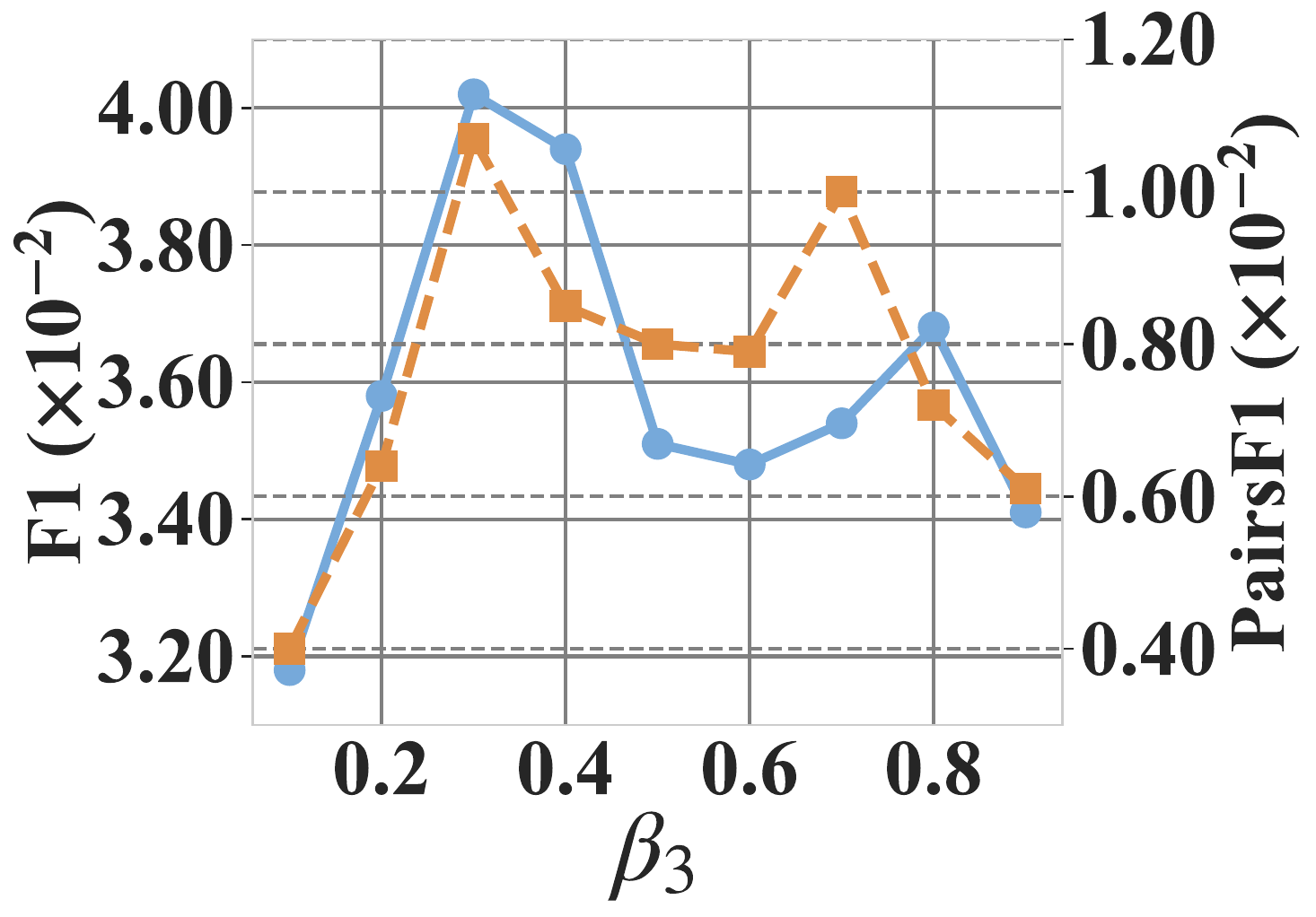}
	\end{subfigure}
	\caption{The effects of discrepancy tolerance parameter ($\sigma_{\tilde{v}^o}$) 
 and various loss function weights ($\beta_1$, $\beta_2$ and $\beta_3$) on the Foursquare dataset w.r.t. the $F_1$ and $PairsF_1$ score.}
	\label{parameter 1}
\end{figure}

\subsection{Hyper-parameter Analysis}
We study how sensitive the framework is to the tolerance parameter $\sigma_{\tilde{v}^o}$ and loss function weights $\beta_1$, $\beta_2$ and $\beta_3$. The results are reported in Fig.~\ref{parameter 1}.
Specifically, $\sigma_{\tilde{v}^o}$ controls the standard deviation of the behavior noise. The results show that both $F_1$ and $PairsF_1$ first show an increasing trend with the increase of $\sigma_{\tilde{v}^o}$ and then decline gradually, where two exceptions are observed when $\sigma_{\tilde{v}^o}$ is set to $0.3$ and $0.4$. This indicates a moderate $\sigma_{\tilde{v}^o}$ (e.g., $0.6$ on Foursquare) would be a good option for SPOT-Trip as it achieves a balance between latent dynamics and true behavior. 

Besides, we assess the effect of each loss weight ($\beta_1$, $\beta_2$, $\beta_3$) by varying one within the range $[0.1, 0.9]$ and setting the other two to $\frac{1 - \beta}{2}$, e.g., varying $\beta_1$ with $\beta_2 = \beta_3 = \frac{1 - \beta_1}{2}$. It can be observed that for the three parameters, the recommendation performance attains its peak when they are set to $0.3$ or $0.4$. This suggests that moderate loss weights help balance the contributions of different loss components during tuning. Based on this observation, we set $\beta_1 = \beta_2 = \beta_3 = 1$ in the final framework, assuming that each loss term is properly normalized to ensure balanced influence. More results regarding parameter sensitivity can be found in Appendix~\ref{more parameter}.

\subsection{The Effect of Data Sparsity}
Tab.~\ref{sparsity} summarizes the performance of various methods trained with limited hometown data. We randomly sample 40\%, 60\%, and 80\% of the hometown trajectories, while preserving the original sequence order and avoiding adjacent duplicate POIs to ensure data quality. It is observed that SPOT-Trip consistently outperforms the strongest baselines, highlighting its robustness in data-scarce scenarios. An interesting observation is that using only 80\% of the hometown data leads to better $F_1$ on Foursquare and $PairsF_1$ on Yelp than using the complete data. We speculate that this improvement may result from the random sampling process, which potentially removes noisy information from the hometown check-ins.

\begin{table}[t]
\caption{The effect of data sparsity. We sample different fractions of the training data.}
\label{sparsity}
\footnotesize
\resizebox{\columnwidth}{!}{%
\begin{tabular}{@{}c|c|cc|cc|cc|cc@{}}
\toprule
\multirow{3}{*}{Dataset} &
  \multirow{3}{*}{Method} &
  \multicolumn{8}{c}{Hometown Data during Training} \\ \cmidrule(l){3-10}
 &
   &
  \multicolumn{2}{c|}{40\%} &
  \multicolumn{2}{c|}{60\%} &
  \multicolumn{2}{c|}{80\%} &
  \multicolumn{2}{c}{100\%} \\
 &
   &
  $F_1(\uparrow)$ & \multicolumn{1}{c|}{$PairsF_1(\uparrow)$} &
  $F_1(\uparrow)$ & \multicolumn{1}{c|}{$PairsF_1(\uparrow)$} &
  $F_1(\uparrow)$ & \multicolumn{1}{c|}{$PairsF_1(\uparrow)$} &
  $F_1(\uparrow)$ & $PairsF_1(\uparrow)$ \\ \midrule
\multirow{3}{*}{Foursquare} &
  Base + KDDC &
  {\ul 0.0325} & \multicolumn{1}{c|}{0.0038} &
  {\ul 0.0348} & \multicolumn{1}{c|}{{\ul 0.0058}} &
  {\ul 0.0364} & \multicolumn{1}{c|}{0.0069} &
  {\ul 0.0375} & 0.0079 \\
 &
  Base + CNN-ODE &
  0.0314 & \multicolumn{1}{c|}{{\ul 0.0046}} &
  0.0336 & \multicolumn{1}{c|}{0.0052} &
  0.0352 & \multicolumn{1}{c|}{{\ul 0.0074}} &
  0.0367 & {\ul 0.0094} \\
 &
  SPOT-Trip &
  \textbf{0.0347} & \multicolumn{1}{c|}{\textbf{0.0073}} &
  \textbf{0.0369} & \multicolumn{1}{c|}{\textbf{0.0085}} &
  \textbf{0.0401} & \multicolumn{1}{c|}{\textbf{0.0091}} &
  \textbf{0.0400} & \textbf{0.0109} \\ \midrule
\multirow{3}{*}{Yelp} &
  Base + KDDC &
  {\ul 0.0321} & \multicolumn{1}{c|}{0.0140} &
  {\ul 0.0331} & \multicolumn{1}{c|}{0.0149} &
  {\ul 0.0338} & \multicolumn{1}{c|}{0.0151} &
  {\ul 0.0341} & 0.0156 \\
 &
  Base + CNN-ODE &
  0.0313 & \multicolumn{1}{c|}{{\ul 0.0154}} &
  0.0327 & \multicolumn{1}{c|}{{\ul 0.0156}} &
  0.0333 & \multicolumn{1}{c|}{{\ul 0.0161}} &
  0.0326 & {\ul 0.0168} \\
 &
  SPOT-Trip &
  \textbf{0.0327} & \multicolumn{1}{c|}{\textbf{0.0162}} &
  \textbf{0.0343} & \multicolumn{1}{c|}{\textbf{0.0178}} &
  \textbf{0.0384} & \multicolumn{1}{c|}{\textbf{0.0216}} &
  \textbf{0.0399} & \textbf{0.0190} \\ \bottomrule
\end{tabular}%
}
\end{table}

\begin{figure}[!htbp]
	\centering
	\begin{subfigure}{0.24\linewidth}
		\centering
		\includegraphics[width=1\linewidth]{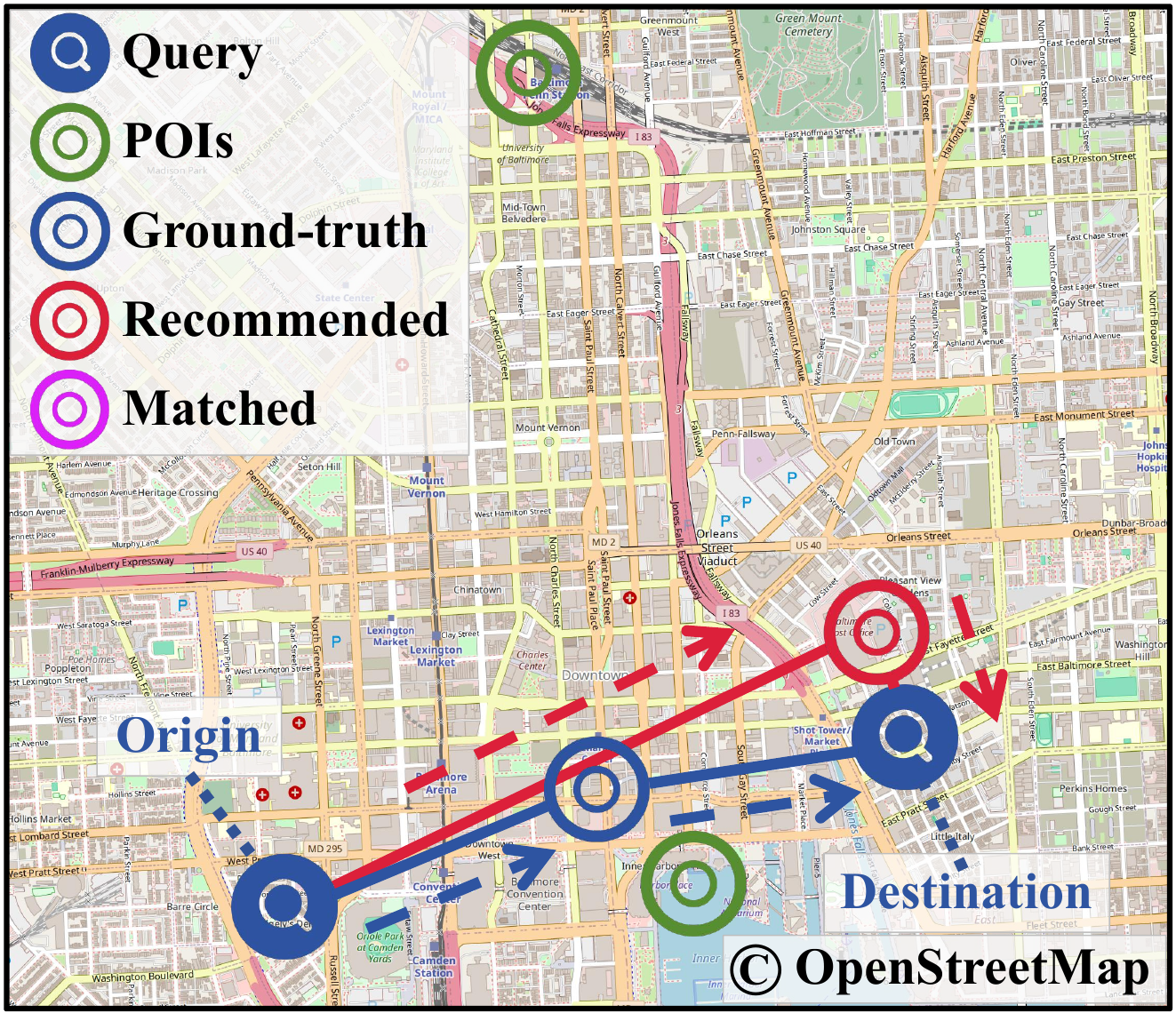}
		\caption{AR-Trip}
	\end{subfigure}
	\centering
	\begin{subfigure}{0.24\linewidth}
		\centering
		\includegraphics[width=1\linewidth]{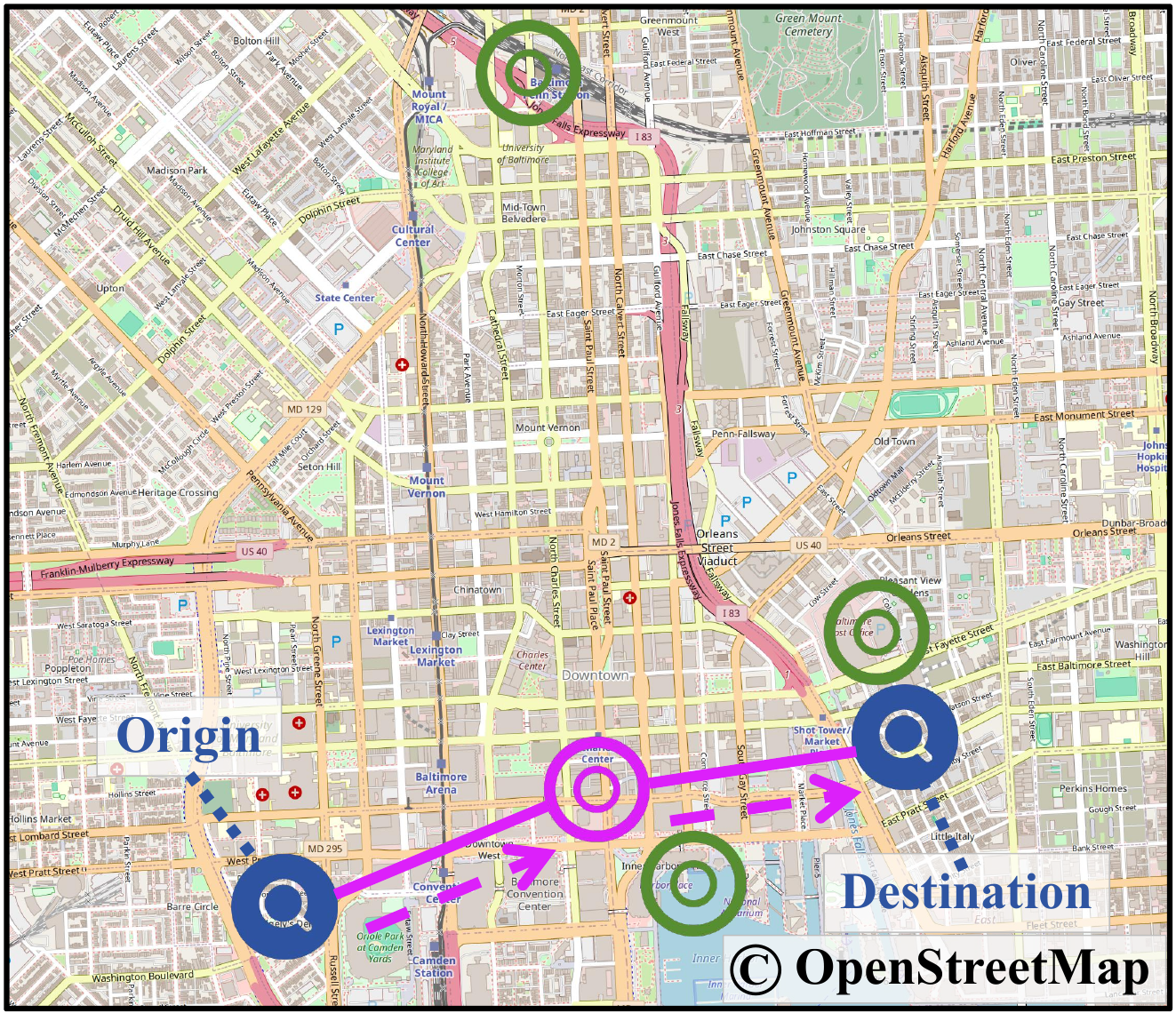}
		\caption{SPOT-Trip}
	\end{subfigure}
	\begin{subfigure}{0.24\linewidth}
		\centering
		\includegraphics[width=1\linewidth]{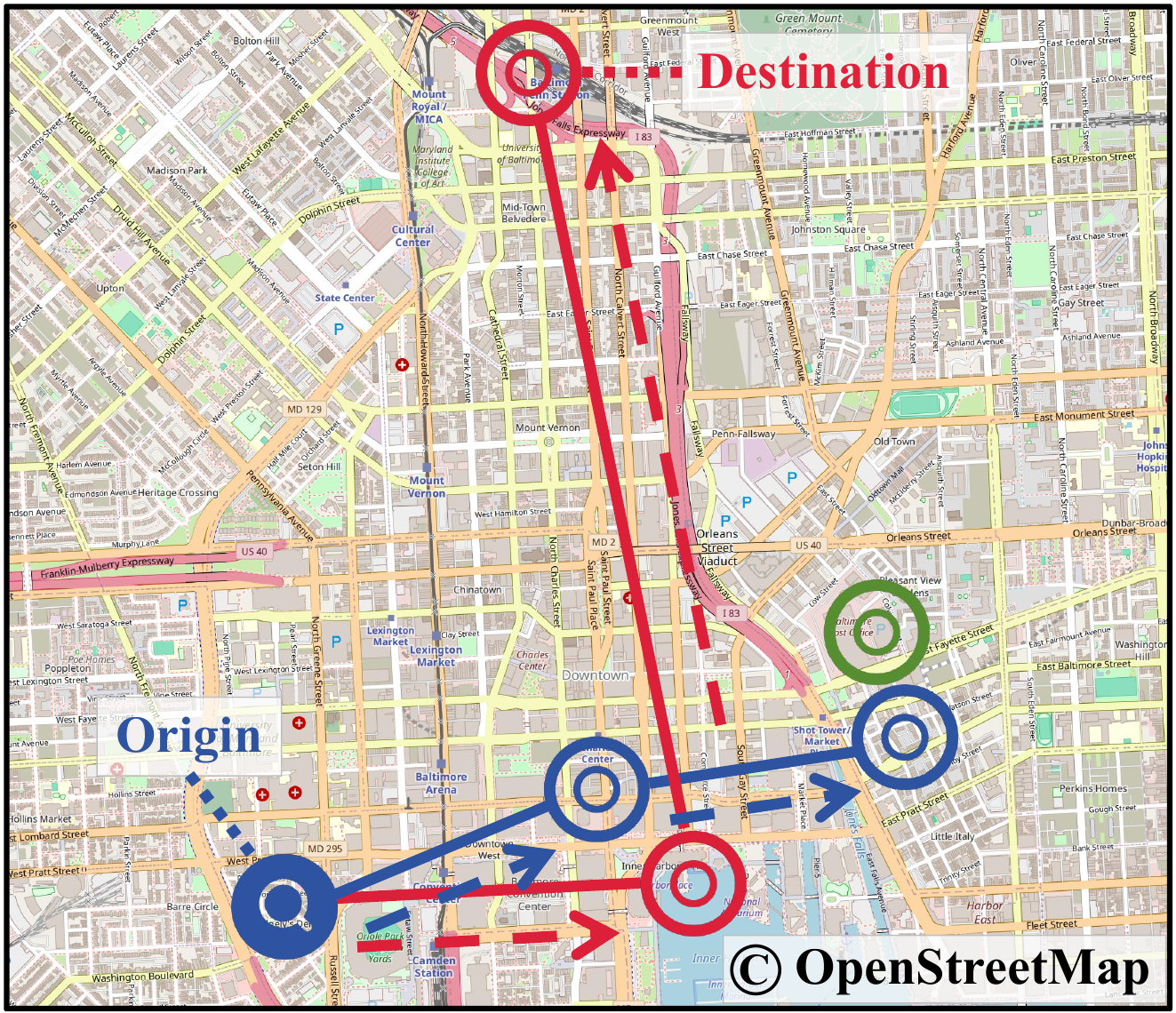}
		\caption{SPOT-Trip (O)}
	\end{subfigure}
	\begin{subfigure}{0.24\linewidth}
		\centering
		\includegraphics[width=1\linewidth]{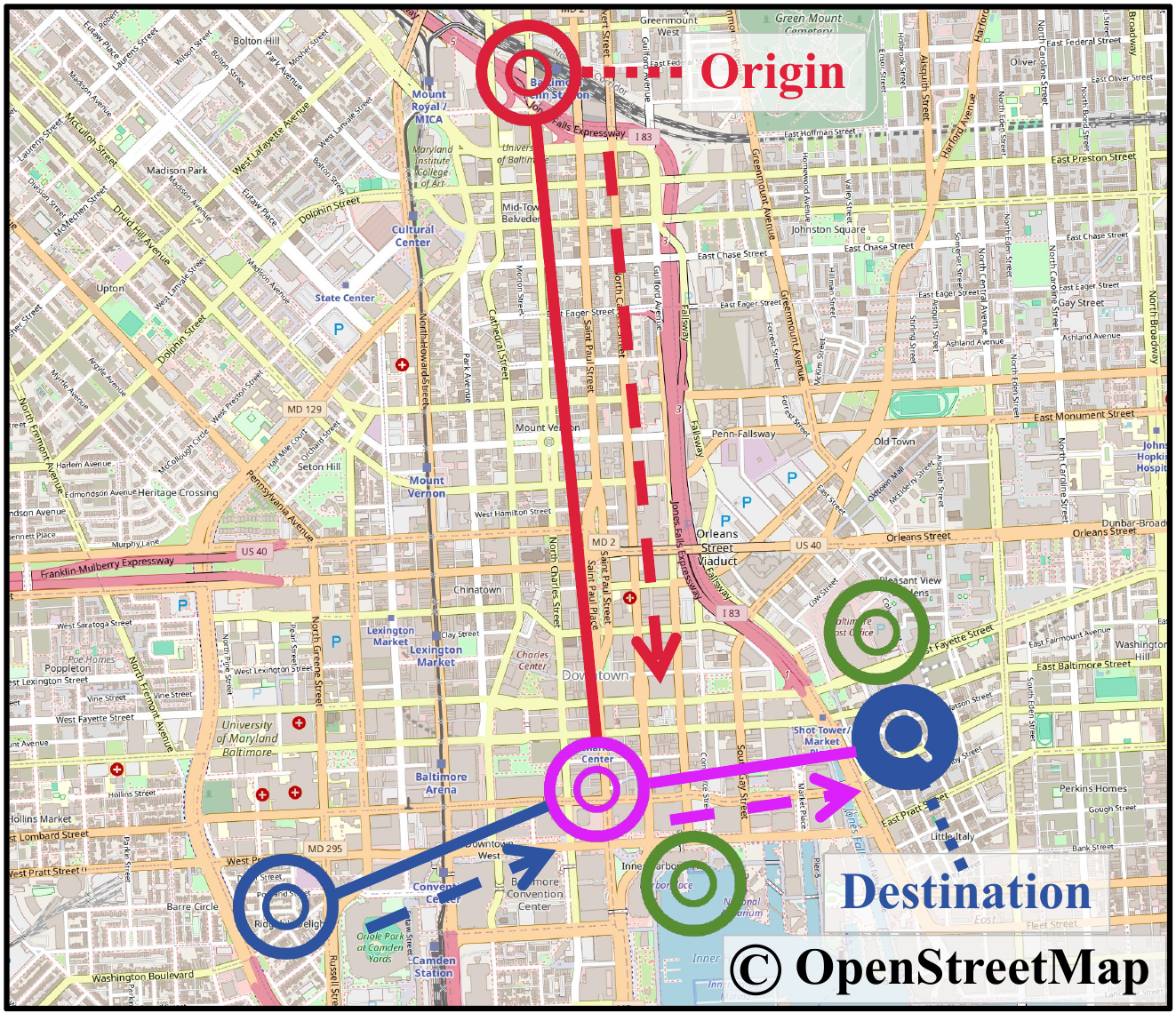}
		\caption{SPOT-Trip (D)}
	\end{subfigure}
	\caption{Visualizations of recommendation results for the user 2964 on Foursquare. (O) denotes a query with a single origin, while (D) denotes a query with a single destination.}
	\label{Case study.}
\end{figure}
\subsection{Case Study}
To intuitively show the effectiveness of SPOT-Trip, we provide a case study on Foursquare as shown in Fig.~\ref{Case study.}. We choose AR-Trip for comparison due to its superior performance. The results show that the recommended intermediate POIs of SPOT-Trip exhibit a remarkable level of alignment with the ground truth. It is clear that SPOT-Trip can accurately trace the right out-of-town trip recommendation. Figs.~\ref{Case study.}(c) and~\ref{Case study.}(d) show that even with only the origin (SPOT-Trip (O)) or destination (SPOT-Trip (D)) provided, SPOT-Trip generates plausible recommendations (e.g., both the predicted destination and origin correspond to train stations), demonstrating its ability to capture realistic user travel patterns. We have provided more case studies in Appendix~\ref{more case}.
\section{Conclusion}
\label{conclusion}
We present SPOT-Trip in this study, a static-dynamic preference aware out-of-town trip recommendation framework, which features three modules. 
First, the knowledge-enhanced static preference learning module constructs a POI attribute knowledge graph and employs relation-aware attention to generate enriched POI embeddings, capturing the static preferences. Second, the ODE-based dynamic preference learning module is proposed to leverage neural ODEs with a temporal point process learning the continuous drift of dynamic user preferences. Together with the static-dynamic preference fusion module, SPOT-Trip delivers superior personalized trips for users traveling from their hometown to unfamiliar regions. Comprehensive experiments on two real datasets offer evidence that SPOT-Trip achieves state-of-the-art accuracy. In the future, an interesting research direction is to apply SPOT-Trip to other spatio-temporal tasks, such as trajectory prediction.

{
\bibliographystyle{abbrv}
\bibliography{main}

\begin{thebibliography}{10}

\bibitem{agrawal2021amortized}
A.~Agrawal and J.~Domke.
\newblock Amortized variational inference for simple hierarchical models.
\newblock {\em NeurIPS}, 34:21388--21399, 2021.

\bibitem{al2024scrollypoi}
I.~Al-Hazwani, T.~Luo, O.~Inel, F.~Ricci, M.~El-Assady, and J.~Bernard.
\newblock Scrollypoi: A narrative-driven interactive recommender system for points-of-interest exploration and explainability.
\newblock In {\em UMAP}, pages 292--304, 2024.

\bibitem{balloccu2022post}
G.~Balloccu, L.~Boratto, G.~Fenu, and M.~Marras.
\newblock Post processing recommender systems with knowledge graphs for recency, popularity, and diversity of explanations.
\newblock In {\em SIGIR}, pages 646--656, 2022.

\bibitem{bordes2013translating}
A.~Bordes, N.~Usunier, A.~Garcia-Duran, J.~Weston, and O.~Yakhnenko.
\newblock Translating embeddings for modeling multi-relational data.
\newblock {\em NeurIPS}, 26, 2013.

\bibitem{cao2024knowledge}
J.~Cao, J.~Fang, Z.~Meng, and S.~Liang.
\newblock Knowledge graph embedding: A survey from the perspective of representation spaces.
\newblock {\em CSUR}, 56(6):1--42, 2024.

\bibitem{chen2016learning}
D.~Chen, C.~S. Ong, and L.~Xie.
\newblock Learning points and routes to recommend trajectories.
\newblock In {\em CIKM}, pages 2227--2232, 2016.

\bibitem{chen2023crom}
P.~Y. Chen, J.~Xiang, D.~H. Cho, Y.~Chang, G.~A. Pershing, H.~T. Maia, M.~M. Chiaramonte, K.~T. Carlberg, and E.~Grinspun.
\newblock Crom: Continuous reduced-order modeling of pdes using implicit neural representations.
\newblock In {\em ICLR}, 2023.

\bibitem{chen2018neural}
R.~T. Chen, Y.~Rubanova, J.~Bettencourt, and D.~K. Duvenaud.
\newblock Neural ordinary differential equations.
\newblock {\em NeurIPS}, 31, 2018.

\bibitem{chen2018neuralode}
R.~T.~Q. Chen, Y.~Rubanova, J.~Bettencourt, and D.~Duvenaud.
\newblock Neural ordinary differential equations.
\newblock {\em NeurIPS}, 2018.

\bibitem{chen2021points}
Y.~Chen, X.~Li, G.~Cong, C.~Long, Z.~Bao, S.~Liu, W.~Gu, and F.~Zhang.
\newblock Points-of-interest relationship inference with spatial-enriched graph neural networks.
\newblock {\em VLDB}, 15(3):504--512, 2021.

\bibitem{choi2021lt}
J.~Choi, J.~Jeon, and N.~Park.
\newblock Lt-ocf: Learnable-time ode-based collaborative filtering.
\newblock In {\em CIKM}, pages 251--260, 2021.

\bibitem{ding2019learning}
J.~Ding, G.~Yu, Y.~Li, D.~Jin, and H.~Gao.
\newblock Learning from hometown and current city: Cross-city poi recommendation via interest drift and transfer learning.
\newblock {\em IMWUT}, 3(4):1--28, 2019.

\bibitem{ference2013location}
G.~Ference, M.~Ye, and W.-C. Lee.
\newblock Location recommendation for out-of-town users in location-based social networks.
\newblock In {\em CIKM}, pages 721--726, 2013.

\bibitem{gao2023dual}
Q.~Gao, W.~Wang, L.~Huang, X.~Yang, T.~Li, and H.~Fujita.
\newblock Dual-grained human mobility learning for location-aware trip recommendation with spatial--temporal graph knowledge fusion.
\newblock {\em Information Fusion}, 92:46--63, 2023.

\bibitem{ghanem2025learning}
P.~Ghanem, A.~Demirkaya, T.~Imbiriba, A.~Ramezani, Z.~Danziger, and D.~Erdogmus.
\newblock Learning physics informed neural odes with partial measurements.
\newblock In {\em AAAI}, volume~39, pages 16799--16807, 2025.

\bibitem{gogleva2022knowledge}
A.~Gogleva, D.~Polychronopoulos, M.~Pfeifer, V.~Poroshin, M.~Ughetto, M.~J. Martin, H.~Thorpe, A.~Bornot, P.~D. Smith, B.~Sidders, et~al.
\newblock Knowledge graph-based recommendation framework identifies drivers of resistance in egfr mutant non-small cell lung cancer.
\newblock {\em Nature Communications}, 13(1):1667, 2022.

\bibitem{gunawan2016orienteering}
A.~Gunawan, H.~C. Lau, and P.~Vansteenwegen.
\newblock Orienteering problem: A survey of recent variants, solution approaches and applications.
\newblock {\em European Journal of Operational Research}, 255(2):315--332, 2016.

\bibitem{he2019joint}
J.~He, J.~Qi, and K.~Ramamohanarao.
\newblock A joint context-aware embedding for trip recommendations.
\newblock In {\em ICDE}, pages 292--303, 2019.

\bibitem{he2016deep}
K.~He, X.~Zhang, S.~Ren, and J.~Sun.
\newblock Deep residual learning for image recognition.
\newblock In {\em ICCV}, pages 770--778, 2016.

\bibitem{Holtzman2020The}
A.~Holtzman, J.~Buys, L.~Du, M.~Forbes, and Y.~Choi.
\newblock The curious case of neural text degeneration.
\newblock In {\em ICLR}, 2020.

\bibitem{iakovlev2025learning}
V.~Iakovlev and H.~L{\"a}hdesm{\"a}ki.
\newblock Learning spatiotemporal dynamical systems from point process observations.
\newblock In {\em ICLR}, 2025.

\bibitem{kong2024kgnext}
X.~Kong, Z.~Chen, J.~Li, J.~Bi, and G.~Shen.
\newblock Kgnext: Knowledge-graph-enhanced transformer for next poi recommendation with uncertain check-ins.
\newblock {\em TCSS}, 2024.

\bibitem{kuo2023bert}
A.-T. Kuo, H.~Chen, and W.-S. Ku.
\newblock Bert-trip: effective and scalable trip representation using attentive contrast learning.
\newblock In {\em ICDE}, pages 612--623. IEEE Computer Society, 2023.

\bibitem{lee2014large}
C.-P. Lee and C.-J. Lin.
\newblock Large-scale linear ranksvm.
\newblock {\em Neural Computation}, 26(4):781--817, 2014.

\bibitem{li2020deep}
D.~Li and Z.~Gong.
\newblock A deep neural network for crossing-city poi recommendations.
\newblock {\em TKDE}, 34(8):3536--3548, 2020.

\bibitem{lin2015learning}
Y.~Lin, Z.~Liu, M.~Sun, Y.~Liu, and X.~Zhu.
\newblock Learning entity and relation embeddings for knowledge graph completion.
\newblock In {\em AAAI}, volume~29, 2015.

\bibitem{liu2024llm}
Q.~Liu, X.~Wu, Y.~Wang, Z.~Zhang, F.~Tian, Y.~Zheng, and X.~Zhao.
\newblock Llm-esr: Large language models enhancement for long-tailed sequential recommendation.
\newblock {\em NeurIPS}, 37:26701--26727, 2024.

\bibitem{liu2024kddc}
Y.~Liu, G.~Shen, C.~Cui, Z.~Zhao, X.~Han, J.~Du, X.~Zhao, and X.~Kong.
\newblock Kddc: Knowledge-driven disentangled causal metric learning for pre-travel out-of-town recommendation.
\newblock In {\em IJCAI}, pages 4--9, 2024.

\bibitem{ludke2023add}
D.~L{\"u}dke, M.~Bilo{\v{s}}, O.~Shchur, M.~Lienen, and S.~G{\"u}nnemann.
\newblock Add and thin: Diffusion for temporal point processes.
\newblock {\em NeurIPS}, 36:56784--56801, 2023.

\bibitem{mercatali2024graph}
G.~Mercatali, A.~Freitas, and J.~Chen.
\newblock Graph neural flows for unveiling systemic interactions among irregularly sampled time series.
\newblock In {\em NeurIPS}, 2024.

\bibitem{mezni2021context}
H.~Mezni, D.~Benslimane, and L.~Bellatreche.
\newblock Context-aware service recommendation based on knowledge graph embedding.
\newblock {\em TKDE}, 34(11):5225--5238, 2021.

\bibitem{qin2024learning}
Y.~Qin, W.~Ju, H.~Wu, X.~Luo, and M.~Zhang.
\newblock Learning graph ode for continuous-time sequential recommendation.
\newblock {\em TKDE}, 36(7):3224--3236, 2024.

\bibitem{sharma2018point}
A.~Sharma, R.~Johnson, F.~Engert, and S.~Linderman.
\newblock Point process latent variable models of larval zebrafish behavior.
\newblock {\em NeurIPS}, 31, 2018.

\bibitem{shu2024analyzing}
W.~Shu, K.~Xu, W.~Tai, T.~Zhong, Y.~Wang, and F.~Zhou.
\newblock Analyzing and mitigating repetitions in trip recommendation.
\newblock In {\em SIGIR}, pages 2276--2280, 2024.

\bibitem{vaswani2017attention}
A.~Vaswani, N.~Shazeer, N.~Parmar, J.~Uszkoreit, L.~Jones, A.~N. Gomez, {\L}.~Kaiser, and I.~Polosukhin.
\newblock Attention is all you need.
\newblock {\em NeurIPS}, 30, 2017.

\bibitem{verma2024climode}
Y.~Verma, M.~Heinonen, and V.~Garg.
\newblock Climode: Climate and weather forecasting with physics-informed neural odes.
\newblock In {\em ICLR}, 2024.

\bibitem{wang2017location}
H.~Wang, Y.~Fu, Q.~Wang, H.~Yin, C.~Du, and H.~Xiong.
\newblock A location-sentiment-aware recommender system for both home-town and out-of-town users.
\newblock In {\em SIGKDD}, pages 1135--1143, 2017.

\bibitem{wei2023lightgt}
Y.~Wei, W.~Liu, F.~Liu, X.~Wang, L.~Nie, and T.-S. Chua.
\newblock Lightgt: A light graph transformer for multimedia recommendation.
\newblock In {\em SIGIR}, pages 1508--1517, 2023.

\bibitem{xin2021out}
H.~Xin, X.~Lu, T.~Xu, H.~Liu, J.~Gu, D.~Dou, and H.~Xiong.
\newblock Out-of-town recommendation with travel intention modeling.
\newblock In {\em AAAI}, volume~35, pages 4529--4536, 2021.

\bibitem{xin2022captor}
H.~Xin, X.~Lu, N.~Zhu, T.~Xu, D.~Dou, and H.~Xiong.
\newblock Captor: A crowd-aware pre-travel recommender system for out-of-town users.
\newblock In {\em SIGIR}, pages 1174--1184, 2022.

\bibitem{xu2024crosspred}
S.~Xu and D.~Guan.
\newblock Crosspred: A cross-city mobility prediction framework for long-distance travelers via poi feature matching.
\newblock In {\em CIKM}, pages 4148--4152, 2024.

\bibitem{xu2020seek}
W.~Xu, S.~Zheng, L.~He, B.~Shao, J.~Yin, and T.-Y. Liu.
\newblock Seek: Segmented embedding of knowledge graphs.
\newblock In {\em ACL}, pages 3888--3897, 2020.

\bibitem{yang2023knowledge}
Y.~Yang, C.~Huang, L.~Xia, and C.~Huang.
\newblock Knowledge graph self-supervised rationalization for recommendation.
\newblock In {\em SIGKDD}, pages 3046--3056, 2023.

\bibitem{yang2022knowledge}
Y.~Yang, C.~Huang, L.~Xia, and C.~Li.
\newblock Knowledge graph contrastive learning for recommendation.
\newblock In {\em SIGIR}, pages 1434--1443, 2022.

\bibitem{yin2023next}
F.~Yin, Y.~Liu, Z.~Shen, L.~Chen, S.~Shang, and P.~Han.
\newblock Next poi recommendation with dynamic graph and explicit dependency.
\newblock In {\em AAAI}, volume~37, pages 4827--4834, 2023.

\bibitem{yin2016joint}
H.~Yin, B.~Cui, X.~Zhou, W.~Wang, Z.~Huang, and S.~Sadiq.
\newblock Joint modeling of user check-in behaviors for real-time point-of-interest recommendation.
\newblock {\em TOIS}, 35(2):1--44, 2016.

\bibitem{zhang2024encoder}
J.~Zhang, M.~Ma, X.~Gao, and G.~Chen.
\newblock Encoder-decoder based route generation model for flexible travel recommendation.
\newblock {\em TSC}, 2024.

\bibitem{zhang2024survey}
Y.~Zhang, X.~Kong, Z.~Shen, J.~Li, Q.~Yi, G.~Shen, and B.~Dong.
\newblock A survey on temporal knowledge graph embedding: Models and applications.
\newblock {\em KBS}, page 112454, 2024.

\end{thebibliography}
}



\appendix

\section{Technical Appendix: Methodology Details}
In this section, we provide more technical details of SPOT-Trip.
\subsection{Alternative Training} \label{Alternative Training}
As discussed in Sec.~\ref{sec:3.1}, the relation-aware knowledge aggregator will be alternately trained with TransE~\cite{bordes2013translating} to improve the multi-relational semantic representation space. The core principle of this translation-based knowledge graph embedding is to ensure that the sum of the head and relation embeddings (i.e., $\mathbf{v}$ and $\mathbf{r}$) approximates the tail embedding $\mathbf{e}$. To quantify similarity, we define $f_d(\cdot)$ as an L1 norm-based measurement function, i.e., $f_d = \|\mathbf{v} + \mathbf{r} - \mathbf{e}\|$. Formally, the translation-based loss $\mathcal{L}_{TransE}$ is expressed as 
\begin{equation}
\mathcal{L}_{TransE} = \sum_{(v, r, e, e') \in \mathcal{G}_K} -\ln \sigma\Big( f_d(\mathbf{v}, \mathbf{r}, \mathbf{e'}) - f_d(\mathbf{v}, \mathbf{r}, \mathbf{e}) \Big), 
\end{equation}
where the negative sample $e'$ is generated by randomly substituting the tail $e$ in the observed triplet $(v, r, e)$ from the knowledge graph $\mathcal{G}_K$. 
\subsection{Approximate Posterior}
\label{Posterior}
We define the approximate posterior with $\psi = [\psi_\mu, \psi_{\sigma^2}]$ as
\begin{equation}\label{eq:approx-post}
  q(\tilde{\mathbf{p}}^o_1; \psi) \;=\; \mathcal{N}\!\bigl(\tilde{\mathbf{p}}^o_1 \,\big|\,
    \psi_\mu,\, \mathrm{diag}\bigl(\psi_{\sigma^2}\bigr)\bigr),
\end{equation}
where $\mathcal{N}$ is the normal distribution, and 
$\mathrm{diag}\bigl(\psi_{\sigma^2}\bigr)$ is a diagonal matrix with
vector $\psi_{\sigma^2}$ on its diagonal. As discussed in Sec.~\ref{sec:3.2}, $\psi$ is obtained from $\tilde{\mathbf{AGG}}$ with the reparameterization sampling, where we further break up $\psi$ into $[\psi_\mu, \psi_{\sigma^2}]$ as follows:
\begin{equation}
  \psi_\mu = \mathrm{Linear}\bigl(\tilde{\mathbf{AGG}}\bigr), 
  \psi_{\sigma^2} = \exp\Bigl(\mathrm{Linear}\bigl(\tilde{\mathbf{AGG}}\bigr)\Bigr),   
\end{equation}
where $\mathrm{Linear}$(·) are separate linear mappings.
\subsection{ELBO}\label{ELBO}
Based on the definitions presented in the previous sections, the ELBO can be formulated as:
\begin{align}
\mathrm{ELBO}=& \int q(\tilde{\mathbf{p}}^o_{1})\,\ln \frac{ p\bigl(\tilde{\mathbf{p}}^o_{1},\{t_n,\tilde{\mathbf{v}}^{o}_n\}_{n=1}^N\bigr)}{ q(\tilde{\mathbf{p}}^o_{1}) }\,d\tilde{\mathbf{p}}^o_{1} 
\nonumber\\  
= &\int q(\tilde{\mathbf{p}}^o_{1})\,\ln \frac{  \prod_{n=1}^{N}p\bigl(\tilde{\mathbf{v}}^{o}_n|\tilde{\mathbf{p}}^o_{1},t_n\bigr)p(\{{t_n}\}_{n-1}^N|\tilde{\mathbf{p}}^o_{1})p(\tilde{\mathbf{p}}^o_{1})}{ q(\tilde{\mathbf{p}}^o_{1}) }\,d\tilde{\mathbf{p}}^o_{1} 
\nonumber \\
=&\int q(\tilde{\mathbf{p}}^o_{1})\,\ln p\bigl(\tilde{\mathbf{p}}^o_{1},\{t_n,\tilde{\mathbf{v}}^{o}_n\}_{n=1}^N\bigr)d\tilde{\mathbf{p}}^o_{1} \nonumber \\
&+ \int q(\tilde{\mathbf{p}}^o_{1})\,\ln p(\{{t_n}\}_{n-1}^N|\tilde{\mathbf{p}}^o_{1}) d\tilde{\mathbf{p}}^o_{1}  \nonumber \\
&+ \int q(\tilde{\mathbf{p}}^o_{1})\,\ln \frac{p(\tilde{\mathbf{p}}^o_{1})}{ q(\tilde{\mathbf{p}}^o_{1}) } d\tilde{\mathbf{p}}^o_{1} \nonumber \\
=&\sum_{n=1}^{N}\mathbb{E}_{q(\tilde{\mathbf{p}}^o_{1})}[\ln p(\tilde{\mathbf{v}}^{o}_n|t_n^o,\tilde{\mathbf{p}}^o_{1})] \nonumber \\
&+ \mathbb{E}_{q(\tilde{\mathbf{p}}^o_{1})}\sum_{n=1}^{N}\ln \lambda(\tilde{\mathbf{p}}^o_{n})-\int\lambda(\tilde{\mathbf{p}}^o)dt  \nonumber \\
&- \mathrm{KL}[q(\tilde{\mathbf{p}}^o_1)||p(\tilde{\mathbf{p}}^o_{1})]
\end{align}
\subsection{Query Representation} \label{query}
A user's query typically reflects her/his inherent semantic preference. Therefore, given the query $Q^o =\{v_0^*,v_N^*,N\}$, we use semantic knowledge aggregation explained in Sec.~\ref{sec:3.1} to obtain the knowledge-enhanced query embeddings $\bar{\mathbf{Q}}^o$. Then we combine $\bar{\mathbf{Q}}^o$ with positional embeddings $\mathbf{N}=\{\mathbf{n}\}_{n=1}^N$ to initialize a Transformer encoder $\mathrm{Trans}_Q$, thereby gaining the position-aware query representation $\mathbf{Q}^o = \mathrm{Trans}_Q(\bar{\mathbf{Q}}^o||\mathbf{N})$. 
\subsection{Optimization and Recommendation Phases}
The detailed algorithm for the optimization and recommendation phases of SPOT-Trip are summarized in Algorithm~\ref{alg:spot-trip 1} and~\ref{alg:spot-trip 2}, to facilitate understanding and implementation. At the beginning of the optimization (Algorithm~\ref{alg:spot-trip 1}), the POI, entity, and relation embedding layers and other parameters in the framework are initialized (line 1). Next, knowledge-aware POI embeddings are obtained via alternating training, while latent POI embeddings are learned through a separate embedding layer (lines 3-4). Next, calculate the static learning loss (lines 6-8), the dynamic learning loss (lines 9-10), and the main recommendation loss (lines 11-12). Through the sum of these three losses (line 13), we can optimize the whole SPOT-Trip. For the recommendation phase (Algorithm~\ref{alg:spot-trip 2}), we first obtain the inferred static preference and dynamic preference for a new user (line 1). Based on these preferences, logits for all POIs within the target region are computed for Top-P sampling (line 2). Finally, the recommended trip is generated according to the user's query (lines 3-9).
\begin{algorithm}[t]
\caption{Optimization phase of SPOT-Trip}
\label{alg:spot-trip 1}
\begin{algorithmic}[1]
\Require POI knowledge graph $\mathcal{G}_k$, user out-of-town trip records $\mathcal{O}$, POI set $\mathcal{V}$, training epochs $E$
\Ensure Out-of-town trip recommender $\mathcal{F}_\theta$ 
\State Initialize POI/entity/relation embedding layers and other framework parameters;
\For{epoch $= 1$ to $E$}
    \State Update POI/entity/relation embedding layers with TransE (Appendix~\ref{Alternative Training}) from $\mathcal{G}_k$;
    \State Get knowledge-aware POI embeddings via Eq.~\ref{eq:1} and latent POI embeddings by $\mathbf{E}(\cdot)$;
    \For{each user $u$ with trip $\xi = (u, \vec{c}_h, \vec{c}_o, a_h, a_o)$}
        \State Get the static preference representations, i.e., $\bar{\mathbf{u}}^h$ and $\bar{\mathbf{u}}^o$;
        \State Obtain inferred out-of-town static preference $\bar{\mathbf{P}}^o$ via Eq.~\ref{eq:2};
        \State Calculate the static learning loss $\mathcal{L}_{S}$ via Eq.~\ref{eq:3};
        \State Obtain inferred out-of-town dynamic preference $\tilde{\mathbf{P}}^o= \{\tilde{\mathbf{p}}^o_n\}_1^N$ via Eq.~\ref{eq:4};
        \State Calculate the dynamic learning loss $\mathcal{L}_{D}$ via Eq.~\ref{eq:9};
        \State Fusion preferences and compute the logits $z_{u,n,i}$ for all POIs located in $a_o$ via Eq.~\ref{eq:10};
        \State Calculate the main recommendation loss $\mathcal{L}_{R}$ via Eq.~\ref{eq:11};
        \State Sum $\mathcal{L}_{S}$, $\mathcal{L}_{D}$ and $\mathcal{L}_{R}$ to obtain $\mathcal{L}$ (Eq.~\ref{eq:12}). Then, update $\mathcal{F}_\theta$ by minimizing $\mathcal{L}$;
    \EndFor
\EndFor
\State \Return $\mathcal{F}_\theta$.
\end{algorithmic}
\end{algorithm}
\begin{algorithm}[t]
\caption{Recommendation phase of SPOT-Trip}
\label{alg:spot-trip 2}
\begin{algorithmic}[1]
\Require User $u^*$, hometown check-ins $\vec{c}_h$, query $Q_o^* = \{v_s^o, v_e^o, N\}$, region $a_o^*$
\Ensure Out-of-town trip $\tau$

\State Obtain the inferred static preference $\bar{\mathbf{P}}^o$ and dynamic preference $\tilde{\mathbf{P}}^o$ via Eqs.~\ref{eq:2} and \ref{eq:4};
\State Fuse preferences and compute the logits $z_{u,n,i}$ for all POIs located in $a_o^*$ via Eq.~\ref{eq:10};
\State Initialize recommended trip $\tau = [v_s^o]$;
\For{$n = 2$ to $N-1$}
    \State Select $v_n^o \sim \text{Top-P}(z_{u,n,i})$;
    \State Append $v_n^o$ to $\tau$;
\EndFor
\State Append $v_N^o = v_e^o$ to $\tau$;
\State \Return $\tau = \{v_1^o, \dots, v_N^o\}$.
\end{algorithmic}
\end{algorithm}

\section{Relate Work}
\label{RelatedWork}
In this section, we will present prior works that are highly relevant to our framework.

\textbf{Out-of-town Recommendation.} Out-of-town recommendation aims to suggest the next likely new POI for users visiting unfamiliar regions. This task is particularly challenging due to issues such as cold start and interest drift. Early work, such as~\cite{ference2013location} addressed the problem by exploring social influence. Subsequent studies~\cite{yin2016joint, wang2017location} primarily adopted latent Dirichlet allocation (LDA) to capture interest drift, incorporating user preferences and POI content. In addition, deep learning-based approaches have been proposed to tackle the problem more effectively. For instance, \cite{xin2021out} leveraged neural topic modeling to conduct a fine-grained analysis of users’ travel intentions, while \cite{xu2024crosspred} exploited cross-city mobility matching to mitigate data sparsity. \cite{xin2022captor} and \cite{liu2024kddc} settled the more intractable pre-travel recommendation with the crowd behavior memory and causal relationship. Despite these efforts, limited attention has been paid to the out-of-town trip recommendation scenario.

\textbf{Trip Recommendation.} Earlier research on trip recommendation predominantly relied on planning-based approaches rooted in the orienteering problem~\cite{gunawan2016orienteering}. For example, Popularity~\cite{chen2016learning} focuses solely on POI popularity and Markov~\cite{chen2016learning} models transitions between POIs. Later methods began to move beyond pure optimization formulations. For instance, C-ILP~\cite{he2019joint} introduces context-aware POI embeddings via linear programming. However, these planning-based methods struggle to capture the complexity and uncertainty of human mobility, motivating the rise of recent learning-based approaches. Transformer-based models like Bert-Trip~\cite{kuo2023bert} treat trip recommendation as a sentence completion task to generate personalized itineraries that align with tourists’ preferences and real-world constraints, while AR-Trip~\cite{shu2024analyzing} leverages a prior position-based matrix to alleviate the issue of repetitive recommendations. While effective to some extent, these methods use only position embeddings to model preference changes and are limited in capturing irregular temporal dynamics, which our proposed SPOT-Trip is designed to overcome.

\textbf{Knowledge Graph-enhanced Recommendation.}
Existing Knowledge Graph (KG)-enhanced approaches for general recommendation tasks can be broadly classified into three categories: embedding-based, path-based, and graph neural network (GNN)-based methods. Embedding-based methods, such as~\cite{mezni2021context}, utilize transition-based models (e.g., TransR~\cite{lin2015learning}) to generate item representations via entity embeddings. Path-based methods~\cite{gogleva2022knowledge, wei2023lightgt} enhance user-item connectivity by constructing meta-paths, yet they heavily rely on domain-specific prior knowledge and manual path design. More recent efforts have shifted towards GNN-based techniques~\cite{balloccu2022post, yang2022knowledge}, which leverage message passing across multi-hop neighbors to capture complex relational structures. KGCL~\cite{yang2022knowledge} and KGRec~\cite{yang2023knowledge} further introduce joint self-supervised learning schemes to mitigate data noise but will increase computational overhead. Therefore, our framework alternately leverages embedding-based and GNN-based methods to enhance the semantic representation of users’ static preferences.

\textbf{Ordinary Differential Equation.} Neural ODEs~\cite{chen2018neuralode} represent a novel framework that extends discrete deep neural networks to continuous-time domains by modeling transformations as ordinary differential equations, effectively generalizing architectures such as ResNet~\cite{he2016deep}. Owing to their strong performance and modeling flexibility, neural ODEs have found wide application across diverse areas, including traffic flow prediction~\cite{mercatali2024graph}, time series forecasting~\cite{verma2024climode}, and continuous dynamical systems~\cite{ghanem2025learning, iakovlev2025learning}. Recently, researchers have begun integrating neural ODEs with GNNs by parameterizing the derivative of hidden node states for sequential recommendation~\cite{choi2021lt, qin2024learning}. In contrast to existing ODE approaches, we introduce neural ODEs to effectively capture the temporal dynamics of user preferences in out-of-town regions.
\section{Additional Experimental Details}
We proceed to provide more details about the experimental settings.
\subsection{Dataset} \label{Dataset details}
The experiments in this study are conducted on two widely used check-in datasets: Foursquare\footnote{https://sites.google.com/site/yangdingqi/home/foursquare-dataset} and Yelp\footnote{https://www.yelp.com.tw/dataset}. Table~\ref{tab:data} presents the statistical summary of these datasets, including details on check-in records and knowledge graph characteristics. For check-in data processing, we first identified users who had check-ins in both their hometown and out-of-town regions. Each user’s check-in sequence was then reformulated into an out-of-town travel record $\xi = (u, \vec{c}_h, \vec{c}_o, a_h, a_o)$ (see \textbf{Definition 3}). To ensure dataset quality, we filtered out Points of Interest (POIs) visited fewer than two times and removed users whose travel records did not meet the following criteria: (1) $\vec{c}_h \geq 4$; (2) $\vec{c}_o \geq 3$; and (3) the frequency of $(a_h, a_o) \geq 10$. Furthermore, we normalized the timestamps and geographic coordinates (latitude and longitude) of all check-ins to the $[0, 1]$ range to facilitate model training. Subsequently, the datasets were randomly partitioned by user into training, validation, and testing sets with a ratio of 80\%, 10\%, and 10\%, respectively. To ensure fairness in evaluation, all users were anonymized. For the construction of knowledge graphs, we generated entity-specific relations using various types of auxiliary information, such as categories, star ratings, user reviews, and associated regions.
\begin{table}[t]
\centering
\caption{Statistics of the two datasets.}
\label{tab:data}
\resizebox{\linewidth}{!}{
\begin{tabular}{@{}c|cccccc|ccc@{}}
\toprule
\multirow{3}{*}{\textbf{Dataset}} & \multicolumn{6}{c|}{\textbf{Raw Records}} & \multicolumn{3}{c}{\textbf{Knowledge Graph}} \\
\cmidrule(lr){2-7} \cmidrule(lr){8-10}
 & \#Users & \#Regions & \#POIs & \#Check-ins & \makecell[c]{\#Hometown\\ Check-ins} & \makecell[c]{\#Out-of-town\\ Check-ins} 
 & \#Relations & \#Entities & \#Triples \\
\midrule
Foursquare & 3,007 & 21 & 23,884 & 126,219 & 109,225 & 16,994 & 2 & 411 & 47,768 \\
Yelp       & 4,417 & 214 & 29,930 & 78,882  & 58,403  & 20,479 & 8 & 53,549 & 353,918 \\
\bottomrule
\end{tabular}}
\end{table}

\subsection{Baseline} \label{Baseline details}
There are two groups of state-of-the-art baselines that are compared within this paper, i.e., 6 trip recommendation-based baselines and 3 out-of-town trip recommendation-based baselines.

\textbf{Trip Recommendation-based Baselines.} These methods primarily focus on modeling POIs within the target out-of-town regions, while paying limited attention to the users’ historical check-in behaviors in their hometowns.
\begin{itemize}[leftmargin=1em]
\item Popularity~\cite{chen2016learning}. It recommends the most popular or frequently visited POIs to a user at each query candidate position.
\item POIRank~\cite{chen2016learning}. It generates a travel trajectory by first ranking POIs using the RankSVM~\cite{lee2014large} method, and then sequentially connecting them based on their ranking scores.
\item GraphTrip~\cite{gao2023dual}. This method proposes a two-stage graph learning framework to model multiple heterogeneous POI graphs, and designs a dual-grained mobility module to capture both coarse-grained category information and fine-grained transitional patterns among POIs.
\item MatTrip~\cite{zhang2024encoder}. This work employs dual long-short-term-memory (LSTM) based encoders to learn users’ category preferences and the geographical proximity of POIs, followed by an attention-based LSTM decoder that generates user-preferred trips with the aid of an optimized search strategy.
\item AR-Trip~\cite{shu2024analyzing}. This approach is based on a Transformer encoder-only architecture and incorporates additional prior position information to recommend trips with low repetition rates.
\item Base. It is a simplified version of our framework that removes all hometown-related information learning and serves as a baseline to examine the contribution of hometown-aware modeling.
\end{itemize}
\textbf{Out-of-town Trip Recommendation-based Baselines.} These baselines distinctively incorporate supplementary historical hometown information for out-of-town recommendations.
\begin{itemize}[leftmargin=1em]
\item Base + KDDC~\cite{liu2024kddc}. KDDC introduces a knowledge graph approach that strengthens semantic interaction and leverages disentangled causal metric learning to align recommended POIs more closely with the target preferences. In the Base + KDDC variant, the knowledge graph approach is integrated to strengthen the Base's static semantic preference alignment ability.
\item Base + CNN-ODE~\cite{iakovlev2025learning}. We isolate the module \textit{ODPL} from SPOT-Trip and replace the Transformer encoder before the Neural ODE with a CNN encoder to support dynamic learning in the Base framework. 
\item Base + PPROC~\cite{iakovlev2025learning}. PPROC combines neural spatiotemporal point processes~\cite{sharma2018point} and neural partial differential equations~\cite{chen2023crom} to improve the prediction of observations at probabilistic locations and timings. It serves as the dynamic preference learning module in the Base + PPROC variant.
\end{itemize}
\subsection{Evaluation Metrics} \label{metrics}
As described in Sec.~\ref{setups}, our evaluation employs four metrics: $F_1$, $PairsF_1$, $Full$-$F_1$, and $Full$-$PairsF_1$. Formally, let $\tau=\{v_1^o,v_2^o,...,v_{N}^o\}$ be the generated trips by a method and $\tau^*$ be the ground truth, the calculation procedures of these metrics are detailed as follows.
\begin{align}
Precision
= \frac{\lvert \tau_{[2:N-1]} \cap \tau_{[2:N-1]}^*\rvert}{\lvert \tau_{[2:N-1]}\rvert},
Recall
= &\frac{\lvert \tau_{[2:N-1]}^* \cap \tau_{[2:N-1]}\rvert}{\lvert \tau_{[2:N-1]}^*\rvert}, \nonumber \\
F_1
= \frac{2\times Precision \times Recall}{Precision + Recall}&.
\end{align}

\begin{align}
Precision_{pairs}
= \frac{N_c}{\displaystyle\binom{\lvert \tau_{[2:N-1]}\rvert}{2}}
\quad,\quad
Recall_{pairs}
= \frac{N_c}{\displaystyle\binom{\lvert \tau_{[2:N-1]}^*\rvert}{2}}, \nonumber \\
PairsF_1
=
\begin{cases}
\dfrac{2\times Precision_{pairs}\times Recall_{pairs}}
      {Precision_{pairs} + Recall_{pairs}}\,, 
& N_c > 0,\\[1em]
0\,, & N_c = 0,
\end{cases}
\end{align}

where $N_c$ is the number of correctly ordered POI-pairs in the recommendation. $\binom{\lvert \tau_{[2:N-1]}}{2}$ and $\binom{\lvert \tau_{[2:N-1]}^*}{2}$ are the total number of ordered pairs, respectively.
Correspondingly, 
\begin{align}
Full\text{-}Precision
= \frac{\lvert \tau \cap \tau^*\rvert}{\lvert \tau\rvert},
Full\text{-}Recall
= \frac{\lvert \tau^* \cap \tau\rvert}{\lvert \tau^*\rvert}, \nonumber \\
Full\text{-}F_1
= \frac{2\times Full\text{-}Precision \times Full\text{-}Recall}{Full\text{-}Precision + Full\text{-}Recall}.
\end{align}

\begin{align}
Full\text{-}Precision_{pairs}
= \frac{N_c}{\displaystyle\binom{\lvert \tau\rvert}{2}}
\quad,\quad
Full\text{-}Recall_{pairs}
= \frac{N_c}{\displaystyle\binom{\lvert \tau^*\rvert}{2}}, \nonumber \\
Full\text{-}PairsF_1
=
\begin{cases}
\dfrac{2\times Full\text{-}Precision_{pairs}\times Full\text{-}Recall_{pairs}}
      {Full\text{-}Precision_{pairs} + Full\text{-}Recall_{pairs}}\,, 
& N_c > 0.\\[1em]
0\,, & N_c = 0.
\end{cases}
\end{align}
Due to the less precise evaluative nature of $Full$-$F_1$ and $Full$-$PairsF_1$, we report them only in the overall comparison experiments.
\subsection{Implementation Details} \label{more implementation details}
We re-implemented the baselines and their hyper-parameters based on the details provided in their original papers and publicly available source codes\footnote{https://github.com/gcooq/GraphTrip}\footnote{https://github.com/Mamingqian/MatTrip}\footnote{https://github.com/Joysmith99/AR-Trip}\footnote{https://github.com/Yinghui-Liu/KDDC}\footnote{https://github.com/yakovlev31/pproc-dyn}. For the two baselines that require periodic information, Graph-Trip~\cite{gao2023dual} and AR-Trip~\cite{shu2024analyzing}, we provided additional hour-of-day features as input. In contrast, PPROC~\cite{iakovlev2025learning} relies on both temporal and spatial points to infer preferences at specific spatiotemporal locations. To support this, we additionally designed two separate gated recurrent unit (GRU) layers to predict the potential time and location points for future trips. Following~\cite{shu2024analyzing}, we fixed the hidden size of all embeddings to $32$ in all our experiments. The optimizer was uniformly chosen as Adam with an initial learning rate of $0.001$ and L2 regularization with a weight of ${10}^{\textnormal{-}5}$. To avoid overfitting, we adopted the early stop strategy with an 8-epoch patience. In addition, the parameters of our framework and all its variants were consistently set as follows: consistent with~\cite{iakovlev2025learning}, the number of Transformer layers in module \textit{ODPL} was set to $4$, while both $f(\cdot)$ and $\lambda(\cdot)$ are implemented as $3$‑layer MLPs. The latent dimensions of these MLPs were searched from $\{16, 32, 64, 128, 256\}$, with $128$ selected as the optimal value based on validation performance. Similarly, we used a differentiable $dopri5$ ODE solver with $rtol$ = $atol$ = $10^{-5}$ from $torchdiffeq$ package~\cite{chen2018neuralode}. For the static-dynamic preference fusion (Sec.~\ref{fusion}), the number of Transformer layers was set to $1$, following the configuration in~\cite{shu2024analyzing}. All Transformer layers employed $4$ attention heads. The hyper-parameter $\sigma_{\tilde{v}^o}$ was tuned separately for each dataset, with the optimal value set to $0.6$ for Foursquare and $0.4$ for Yelp. In the optimization stage, the weights of the loss terms $\beta_1$, $\beta_2$ and $\beta_3$ for two datasets were set as $1$.

\section{Additional Experimental Results}
In this section, we will show more experimental results to further analyze the effectiveness of our SPOT-Trip.
\subsection{Ablation Study} \label{more ablation}
\begin{figure*}
  \centering
  \includegraphics[width=\linewidth]{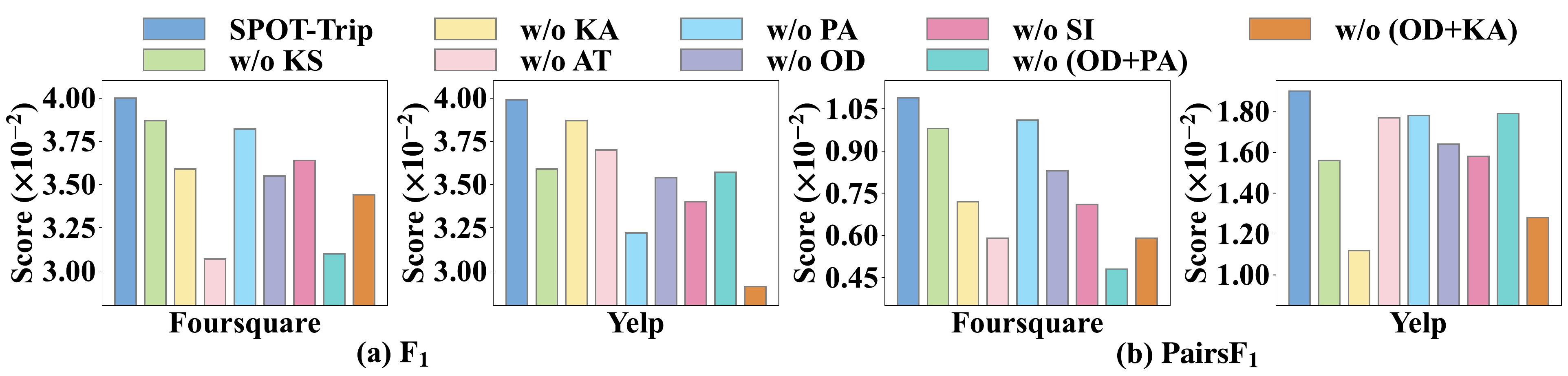}
  \caption{Performance of SPOT-Trip and its variants on two datasets.}
  \label{Ablation 1 all}
\end{figure*}
\begin{table}[]
\setlength\tabcolsep{13pt}
\caption{Several Component Replacement Ablation Experiments.}
\label{replacement ablation}
\centering
\small
\begin{tabular}{@{}ccccc@{}}
\toprule
\multicolumn{1}{c|}{\multirow{2}{*}{Method}}       & \multicolumn{2}{c|}{Foursquare}                        & \multicolumn{2}{c}{Yelp}          \\ \cmidrule(l){2-5} 
\multicolumn{1}{c|}{}                   & $F_1$  & \multicolumn{1}{c|}{$PairsF_1$} & $F_1$  & $PairsF_1$ \\ \midrule
\multicolumn{1}{c|}{SPOT-Trip}                     & \textbf{0.0400} & \multicolumn{1}{c|}{\textbf{0.0109}} & \textbf{0.0399} & \textbf{0.0190} \\ \midrule
\multicolumn{5}{c}{The replacements of the two main modules}                                           \\ \midrule
\multicolumn{1}{c|}{Base + KDDC~\cite{liu2024kddc} + \textit{ODPL}}    & 0.0385 & \multicolumn{1}{c|}{0.0089}    & 0.0361 & 0.0174    \\
\multicolumn{1}{c|}{Base + \textit{KSPL} + CNN-ODE~\cite{iakovlev2025learning}} & 0.0390 & \multicolumn{1}{c|}{0.0107}    & 0.0351 & 0.0186    \\ \midrule
\multicolumn{5}{c}{The replacements of the alternative training method}                                \\ \midrule
\multicolumn{1}{c|}{TransR~\cite{lin2015learning}}             & 0.0378 & \multicolumn{1}{c|}{0.0091}    & 0.0366 & 0.0182    \\
\multicolumn{1}{c|}{SEEK ($2$)~\cite{xu2020seek}}           & 0.0348 & \multicolumn{1}{c|}{0.0082}    & 0.0342 & 0.0146    \\
\multicolumn{1}{c|}{SEEK ($4$)~\cite{xu2020seek}}           & 0.0376 & \multicolumn{1}{c|}{0.0087}    & 0.0346 & 0.0154    \\
\multicolumn{1}{c|}{SEEK ($8$)~\cite{xu2020seek}}           & 0.0394 & \multicolumn{1}{c|}{0.0108}    & 0.0361 & 0.0183    \\
\multicolumn{1}{c|}{SEEK ($16$)~\cite{xu2020seek}}          & 0.0346 & \multicolumn{1}{c|}{0.0073}    & 0.0335 & 0.0134    \\ \midrule
\multicolumn{5}{c}{The replacement of the parallel design}                                \\ \midrule
\multicolumn{1}{c|}{SPOT-Trip (\textit{ODPL} -> \textit{KSPL})} & 0.0351          & \multicolumn{1}{c|}{0.0078}          & 0.0344          & 0.0153          \\ \bottomrule
\end{tabular}%
\end{table}
Beyond the three variants introduced in Sec.~\ref{ablation} (\textbf{w/o KS}, \textbf{w/o OD}, and \textbf{w/o SI}), Fig.~\ref{Ablation 1 all} presents a more comprehensive ablation study involving the removal of various modules and submodules. These include: (1) \textbf{w/o KA}. SPOT-Trip removes the semantic knowledge aggregation in \textit{KSPL}; (2) \textbf{w/o AT}. SPOT-Trip removes the alternative training in KA; (3) \textbf{w/o PA}. SPOT-Trip removes static preference alignment in \textit{KSPL}; and two combination removals: (4) \textbf{w/o (OD+PA)}. SPOT-Trip makes recommendations solely based on the knowledge-enhanced query embedding; and (5) \textbf{w/o (OD+KA)}. SPOT-Trip utilizes the raw hometown representation, without semantic enhancement, to support the recommendation. 
Overall, the performance degradation observed in each variant demonstrates the effectiveness of every component within the SPOT-Trip architecture.
In terms of $F_1$ score, the most pronounced performance drop was observed for the \textbf{w/o AT} variant on Foursquare and the \textbf{w/o (OD+KA)} variant on Yelp. This discrepancy may stem from dataset scale: Foursquare’s sparse triples require additional training, whereas Yelp’s larger dataset risks noise without sufficient semantic augmentation. Regarding the PairsF1 score, the \textbf{w/o (OD+PA)} ablation on Foursquare and the \textbf{w/o SM} ablation on Yelp exhibit the largest performance drops. The findings demonstrate the crucial importance of learning static and dynamic preferences. 

In addition to the removal-based ablation studies, we further conducted a series of component replacement ablation experiments in Tab.~\ref{replacement ablation}.
Firstly, we replace the two main modules of our framework with top-performing baselines: KDDC (Base + KDDC + \textit{ODPL}) and CNN-ODE (Base + \textit{KSPL} + CNN-ODE), to further assess the relative contribution of each component. Compared to our full framework, the variant Base + KDDC + \textit{ODPL} results in a 22.47\% drop in $PairsF_1$ on the Foursquare dataset, while Base + \textit{KSPL} + CNN-ODE leads to a 13.68\% decrease in $F_1$ on Yelp. These results highlight the unique advantages and effectiveness of our framework design. Secondly, to assess the effectiveness of our alternating training strategy, we replace TransE in \textit{KSPL} with TransR~\cite{lin2015learning} and SEEK~\cite{xu2020seek}. TransR models entities and relations in separate vector spaces to capture complex relational patterns, while SEEK introduces segmented embeddings and interaction-aware scoring functions to enhance expressiveness. For SEEK ($k$), we set the number of embedding segments $k$ to $\{2, 4, 8, 16\}$. The observed performance drops indicate that TransE is better suited for our framework, likely due to its low model complexity and stable optimization behavior under alternating training. In contrast, more expressive models such as TransR and SEEK may require more careful tuning to fully realize their potential. Finally, we replace the original parallel architecture with a sequential variant, where the knowledge-enhanced POI embeddings are directly used for modeling users' dynamic preferences, denoted as SPOT-Trip (\textit{ODPL} -> \textit{KSPL}). This cascaded design significantly underperforms our parallel architecture, likely due to interference between static and dynamic preference signals when modeled in a non-decoupled manner. This result further validates the advantage of our parallel modeling design.
\subsection{Hyper-parameter Analysis} \label{more parameter}
\begin{figure}[t]
	\centering
	\begin{subfigure}{0.245\linewidth}
		\centering
		\includegraphics[width=1\linewidth]{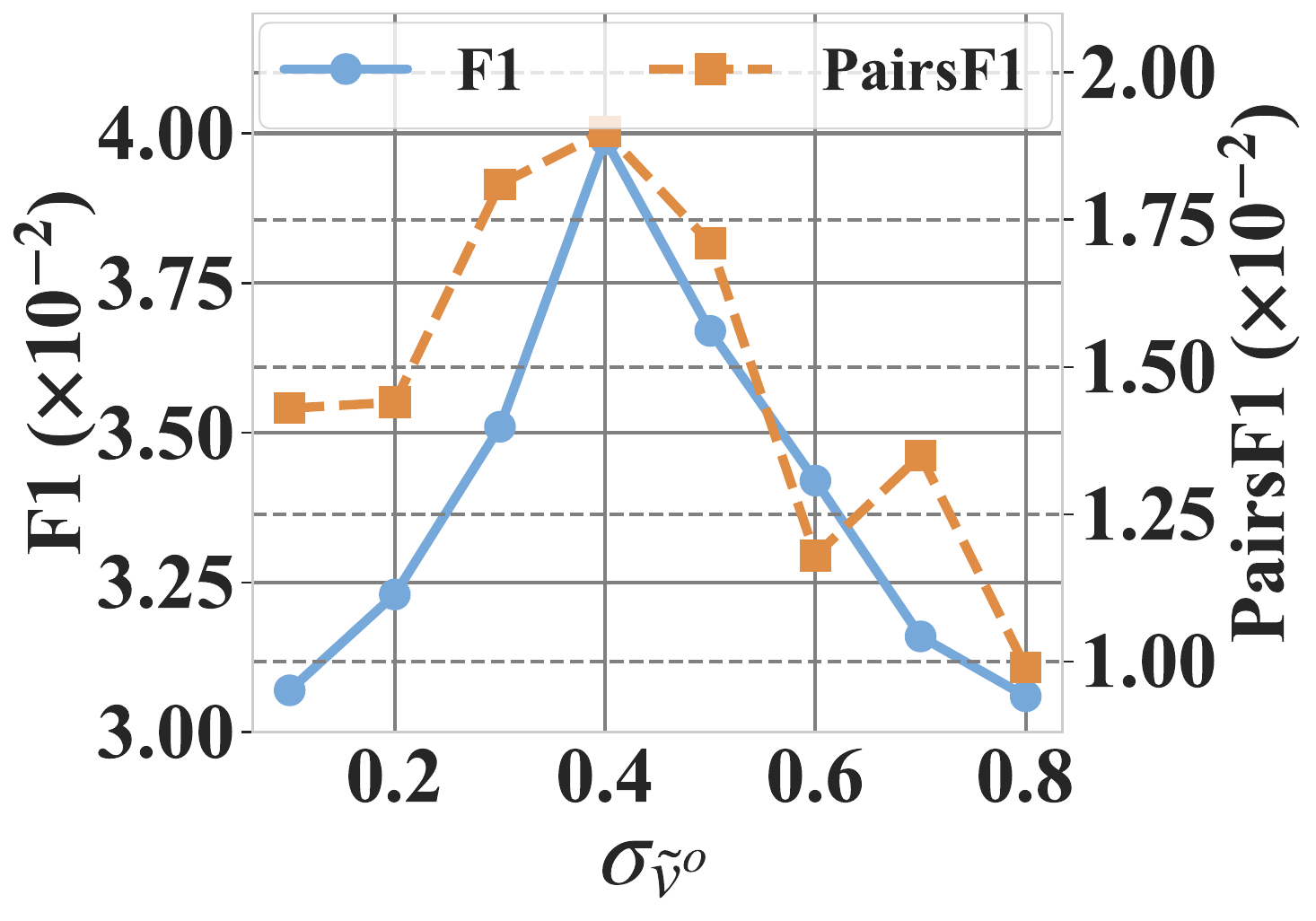}
	\end{subfigure}
	\centering
	\begin{subfigure}{0.245\linewidth}
		\centering
		\includegraphics[width=1\linewidth]{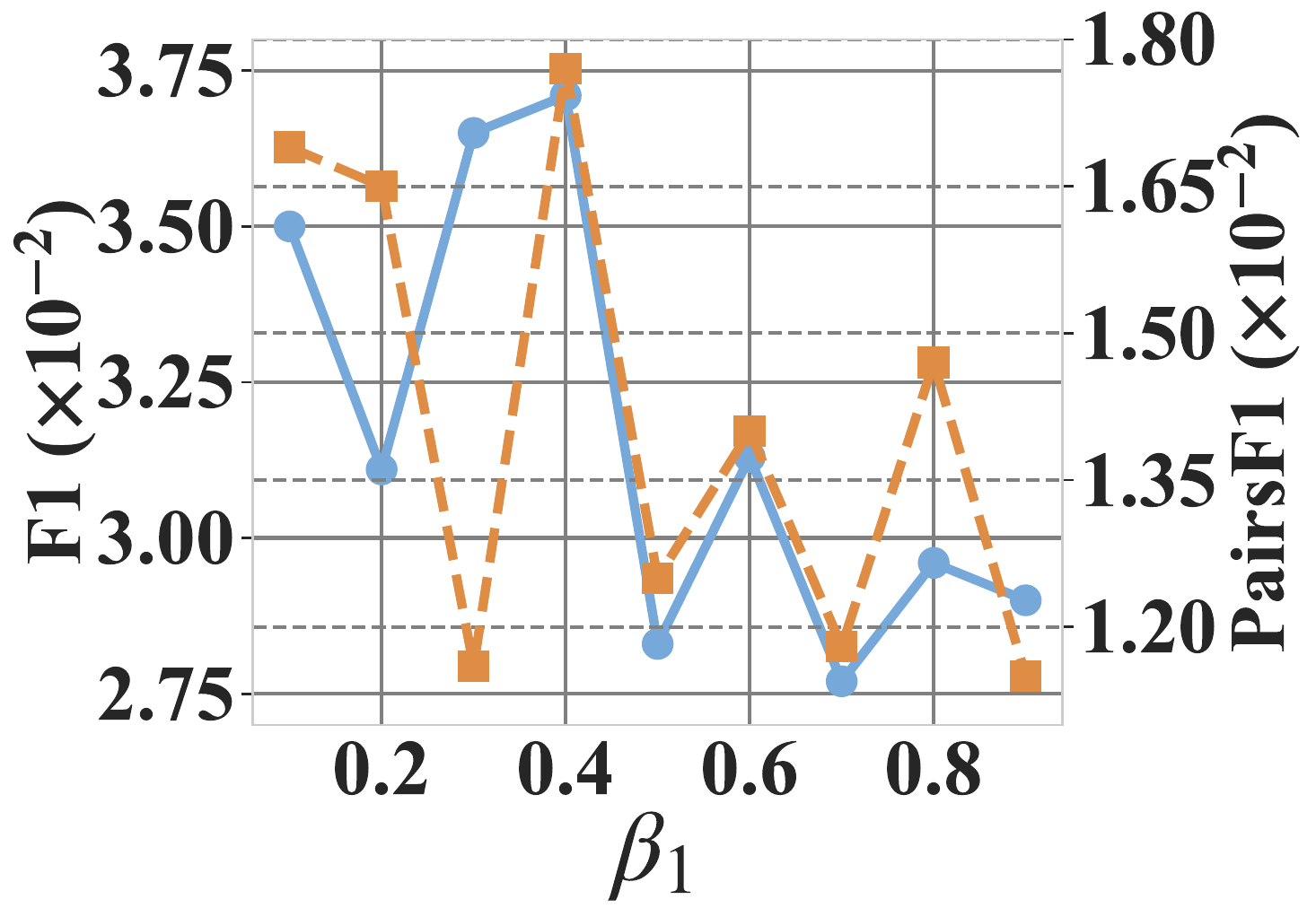}
	\end{subfigure}
	\begin{subfigure}{0.245\linewidth}
		\centering
		\includegraphics[width=1\linewidth]{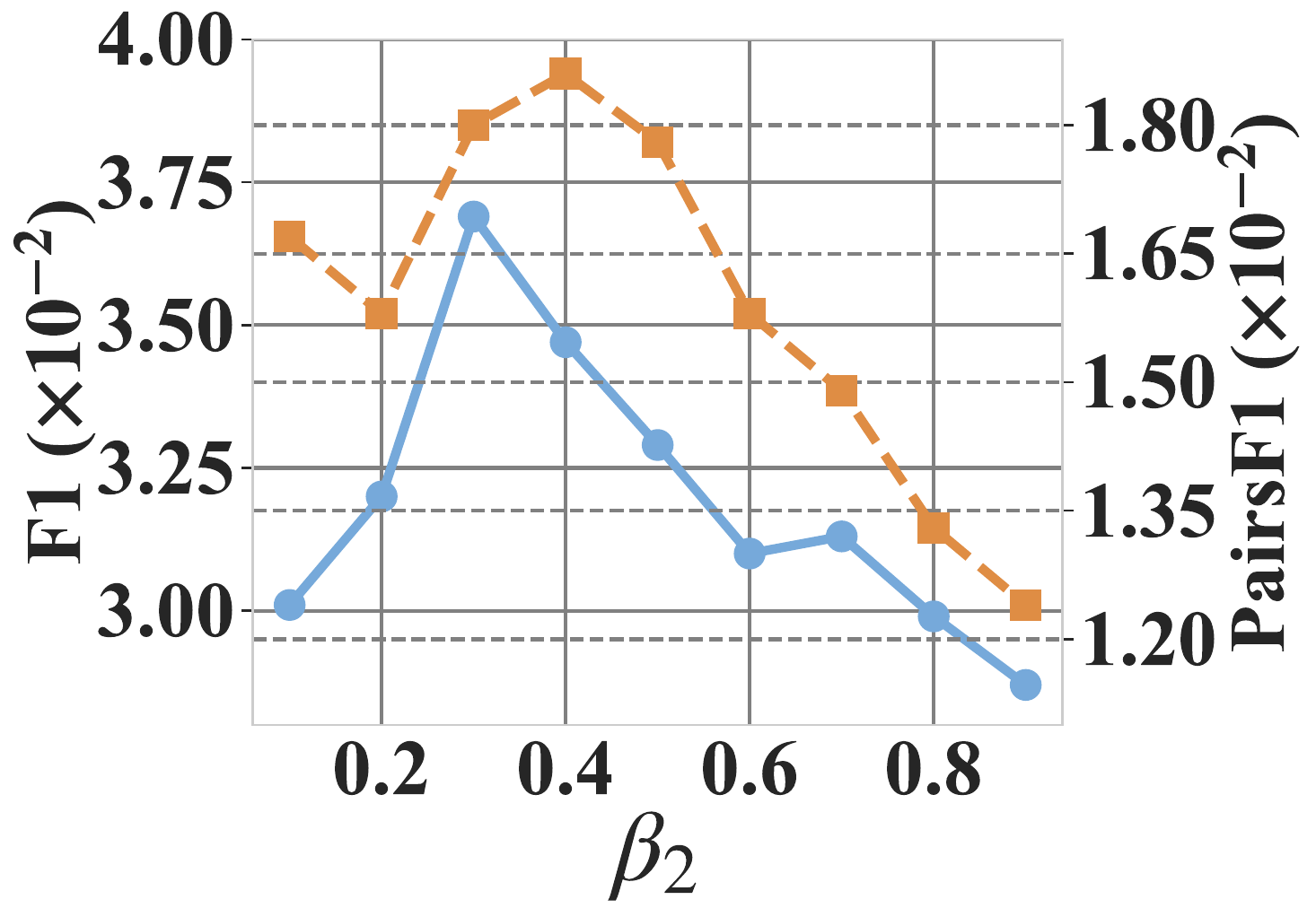}
	\end{subfigure}
	\begin{subfigure}{0.245\linewidth}
		\centering
		\includegraphics[width=1\linewidth]{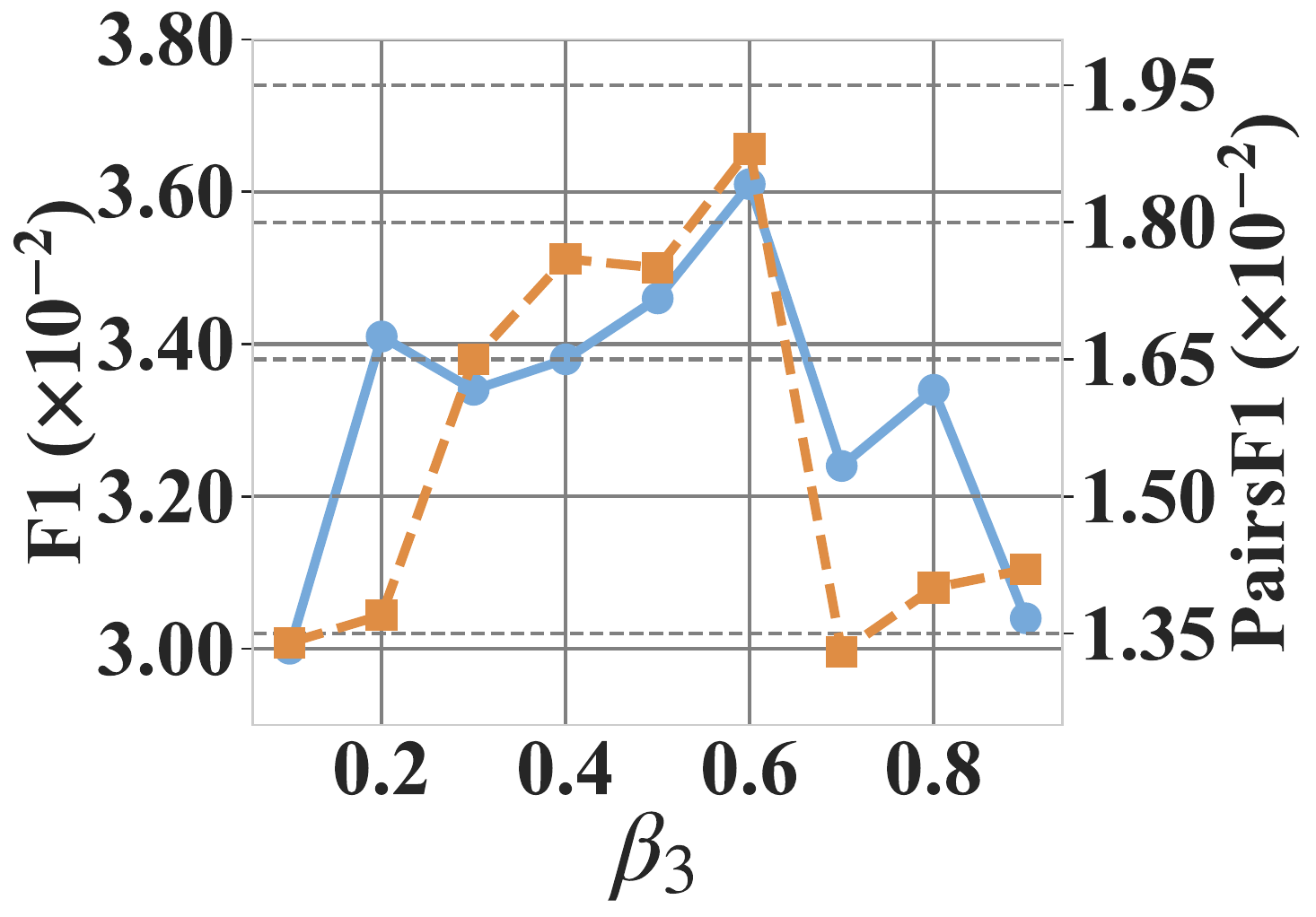}
	\end{subfigure}
	\caption{The effects of discrepancy tolerance parameter ($\sigma_{\tilde{v}^o}$) 
 and various loss function weights ($\beta_1$, $\beta_2$ and $\beta_3$) on the Yelp dataset w.r.t. the $F_1$ and $PairsF_1$ score.}
	\label{parameter 2}
\end{figure}
\begin{figure}[t]
  \centering
  \setlength{\tabcolsep}{0pt}
\begin{tabular}{@{}cccc@{}}
\includegraphics[width=0.245\linewidth]{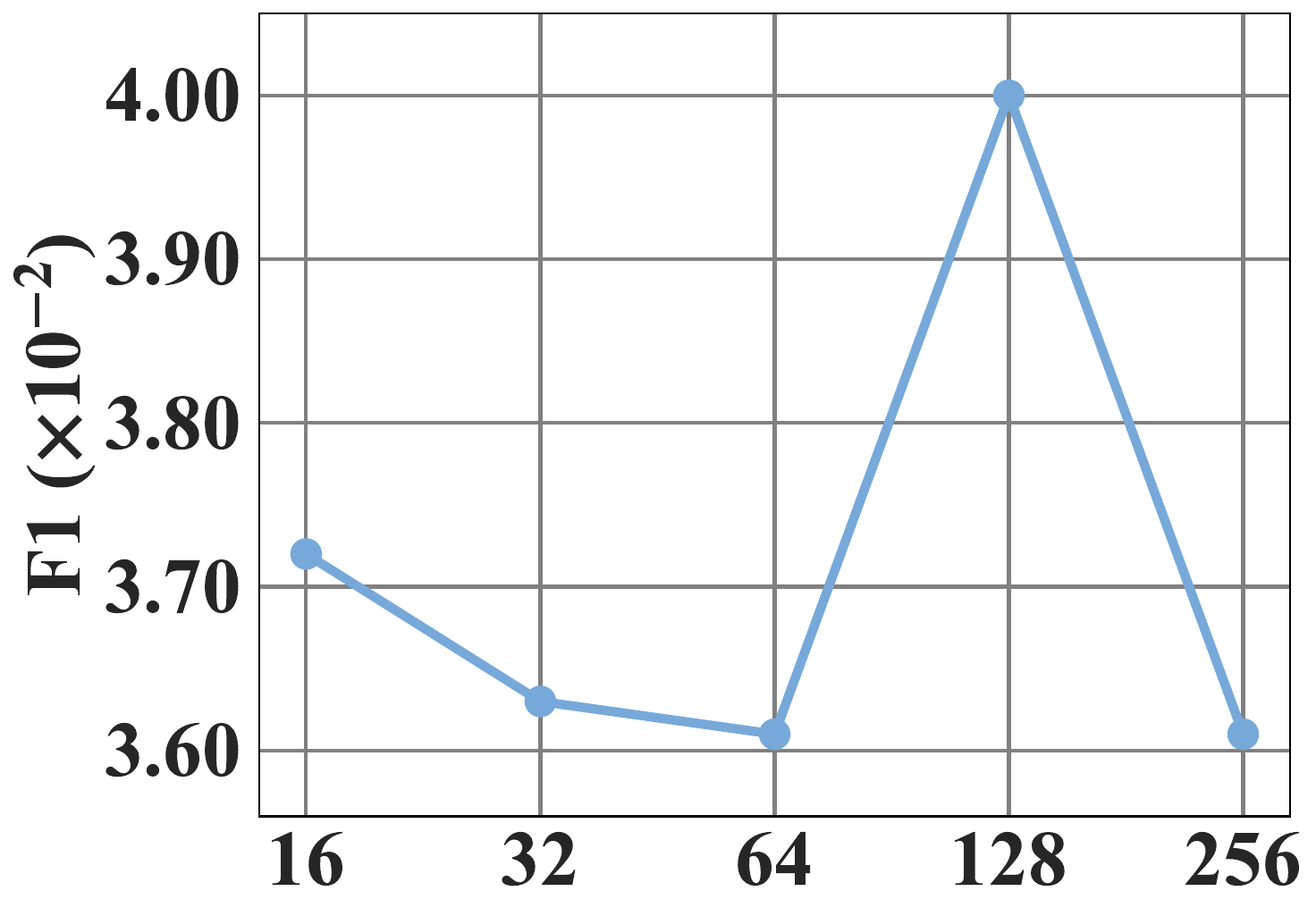} &
\includegraphics[width=0.245\linewidth]{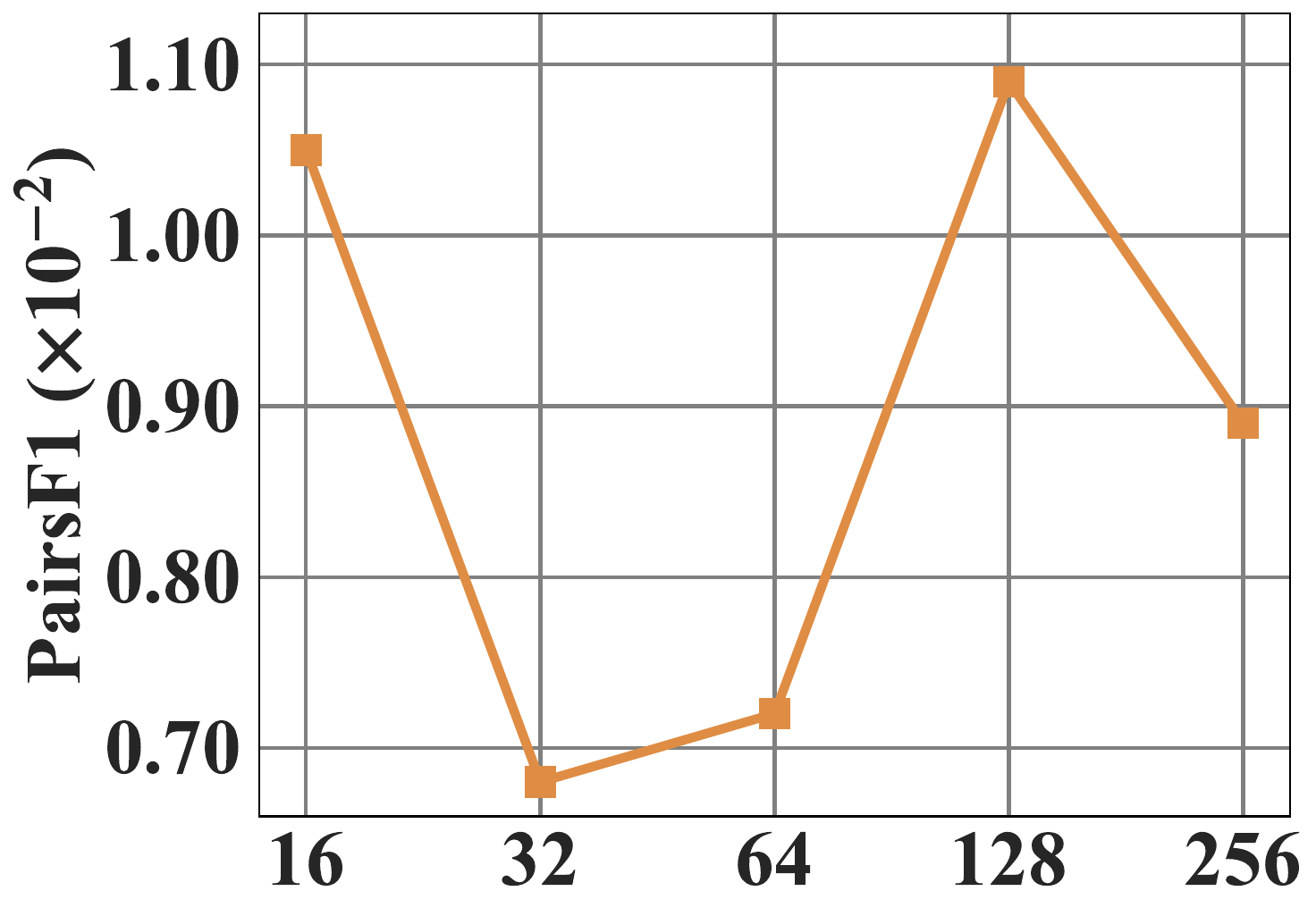} &
\includegraphics[width=0.245\linewidth]{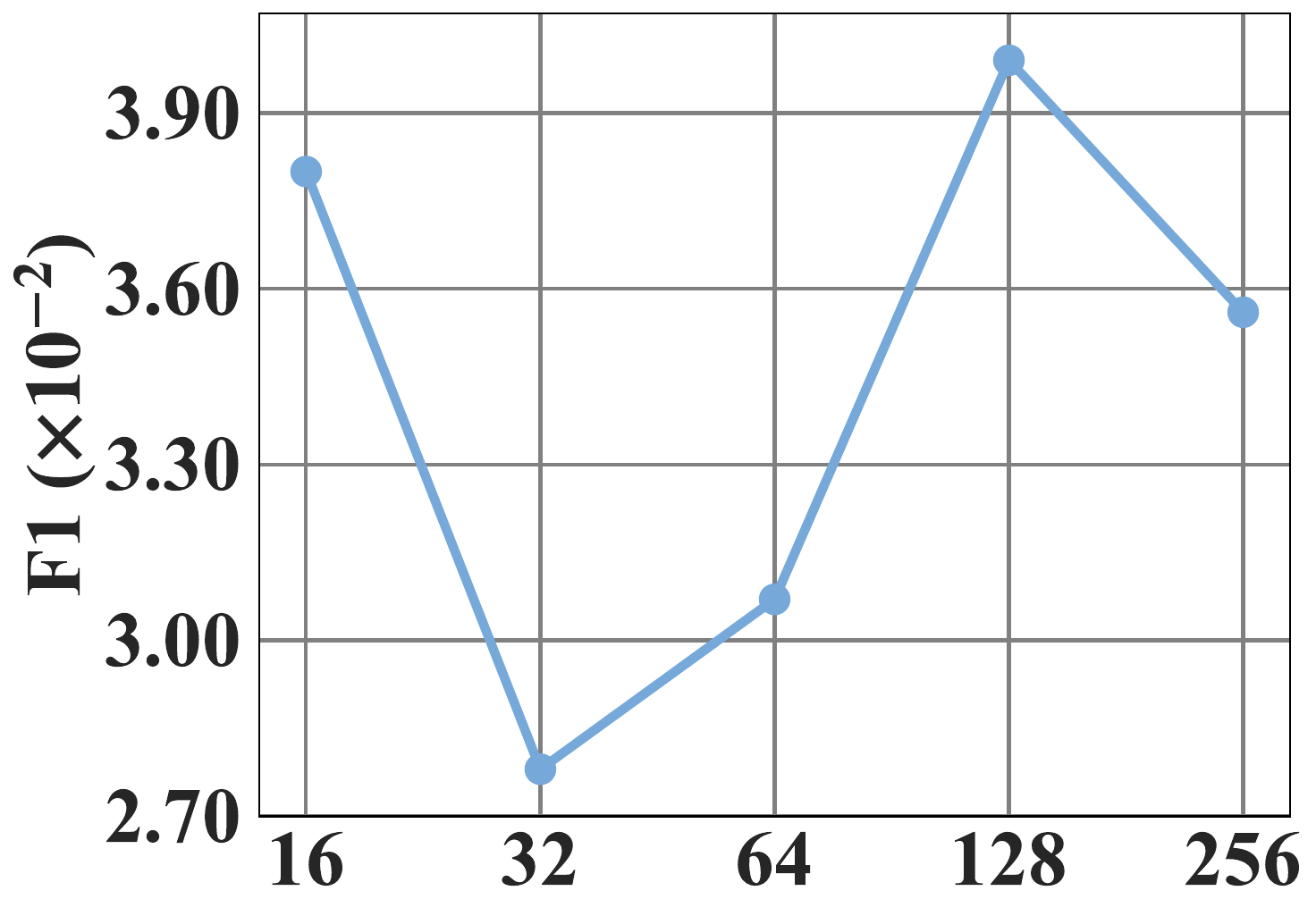} &
\includegraphics[width=0.245\linewidth]{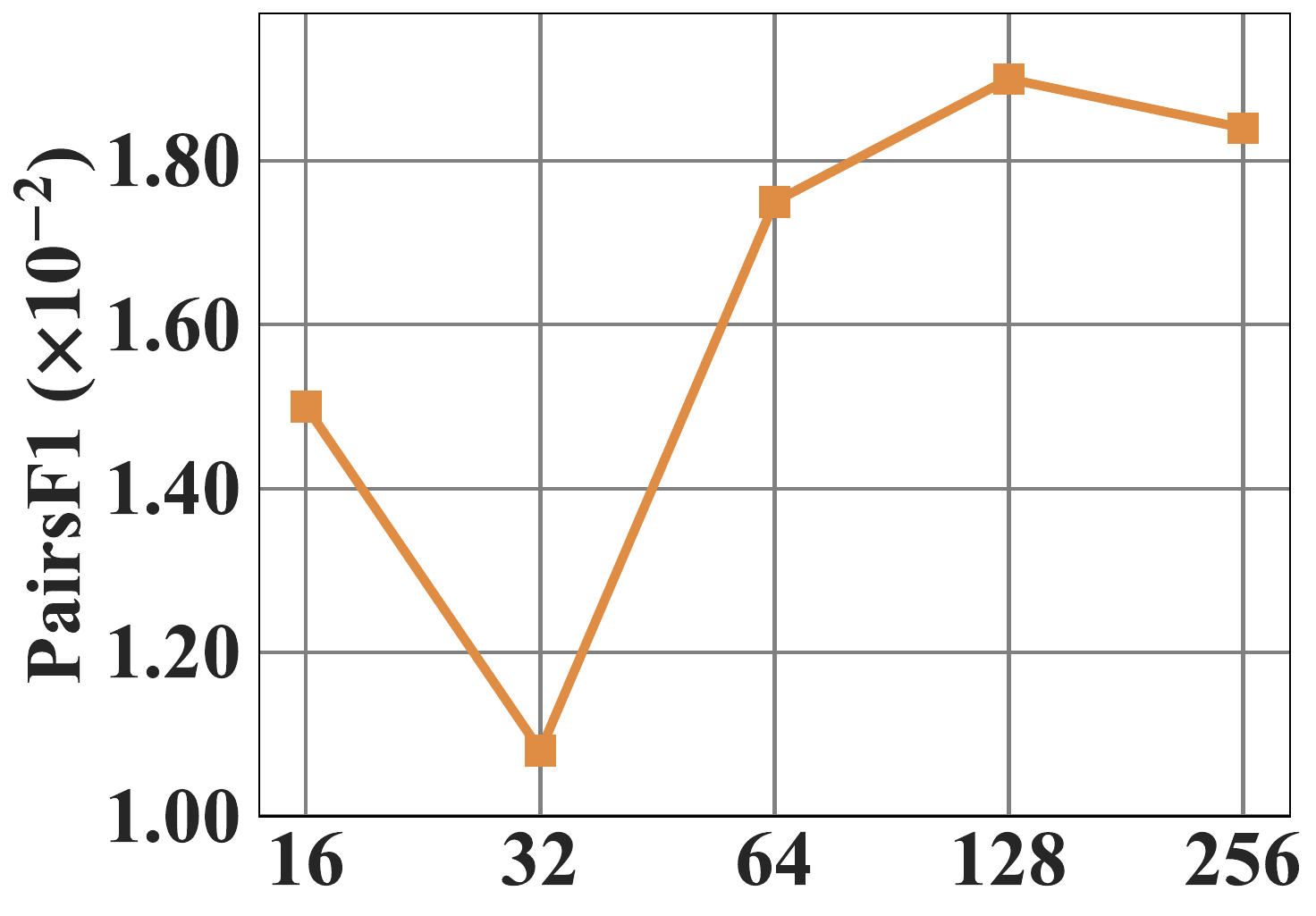} \\[-0.2em]
    \multicolumn{2}{c}{\fontsize{8}{9}\selectfont  (a) Foursquare} &
    \multicolumn{2}{c}{\fontsize{8}{9}\selectfont  (b) Yelp} \\
  \end{tabular}
\caption{The effects of the number of hidden states of MLPs in both $f$(·) and $\lambda$(·) on two datasets.}
\label{parameter 3}
\end{figure}
Fig.~\ref{parameter 2} illustrates how varying the tolerance parameter $\sigma_{\tilde{v}^o}$ and the loss function weights $\beta_1$, $\beta_2$ and $\beta_3$ affects the performance of SPOT-Trip on the Yelp dataset. This line chart result of $\sigma_{\tilde{v}^o}$ shows a similar trend to the result in Fig.~\ref{parameter 1}: as $\sigma_{\tilde{v}^o}$ increases, both $F_1$ and $PairsF_1$ scores rise to a maximum at $0.4$ and then gradually fall, with only a minor rebound at $0.7$. Therefore, setting $\sigma_{\tilde{v}^o} = 0.4$ balances model expressiveness with fidelity to real user behavior. For $\beta_1$, $\beta_2$ and $\beta_3$, we observe that $F_1$ and $PairsF_1$ generally peak when $\beta = 0.3$ or $0.4$. The only notable exceptions occur at $\beta_1 = 0.3$ and $\beta_3 = 0.6$, where performance slightly deviates. Nevertheless, these results confirm that moderate weighting yields the best balance across loss terms. Accordingly, we retain $\beta_1 = \beta_2 = \beta_3$ = 1 in our final framework, assuming proper normalization of each component.

Moreover, we simultaneously vary the latent dimensions of the ODE functions $f$(·) and $\lambda$(·) from $16$ to $256$ to study their joint impact on user dynamic preference modeling in SPOT-Trip. As shown in Fig.~\ref{parameter 3}, we vary the latent dimensions from $16$ to $256$ and evaluate the performance on both Foursquare and Yelp datasets using $F_1$ and $PairsF_1$ as metrics. On Foursquare, both $F_1$ and $PairsF_1$ initially decrease at smaller dimensions ($32$ and $64$), but sharply increase at $128$, which yields the best performance. A slight degradation is observed at $256$, likely due to overfitting or increased optimization difficulty. A similar trend is observed on Yelp, where performance significantly improves with larger dimensions and peaks at $128$ before plateauing or dropping slightly at $256$. These findings suggest that the choice of latent dimension also plays a critical role in Neural ODE-based preference modeling. Extremely small dimensions may limit expressiveness, while overly large dimensions can introduce instability. Thus, a latent dimension of $128$ is adopted in our framework.
\subsection{Case Study} \label{more case}
\begin{figure}[t]
	\centering

	\begin{subfigure}{0.245\linewidth}
		\centering
		\includegraphics[width=1\linewidth]{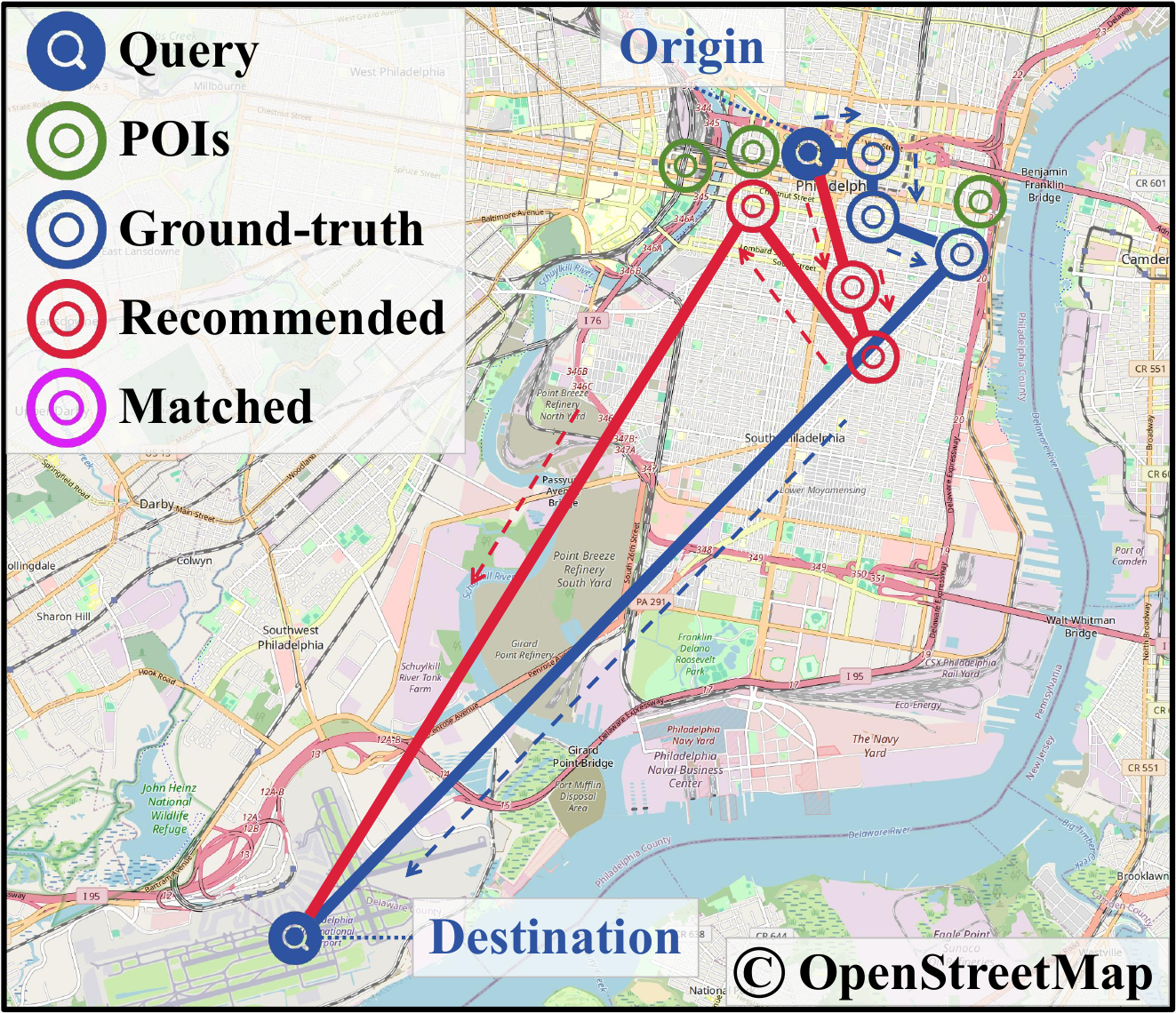}
		\caption{AR-Trip}
	\end{subfigure}
	\begin{subfigure}{0.245\linewidth}
		\centering
		\includegraphics[width=1\linewidth]{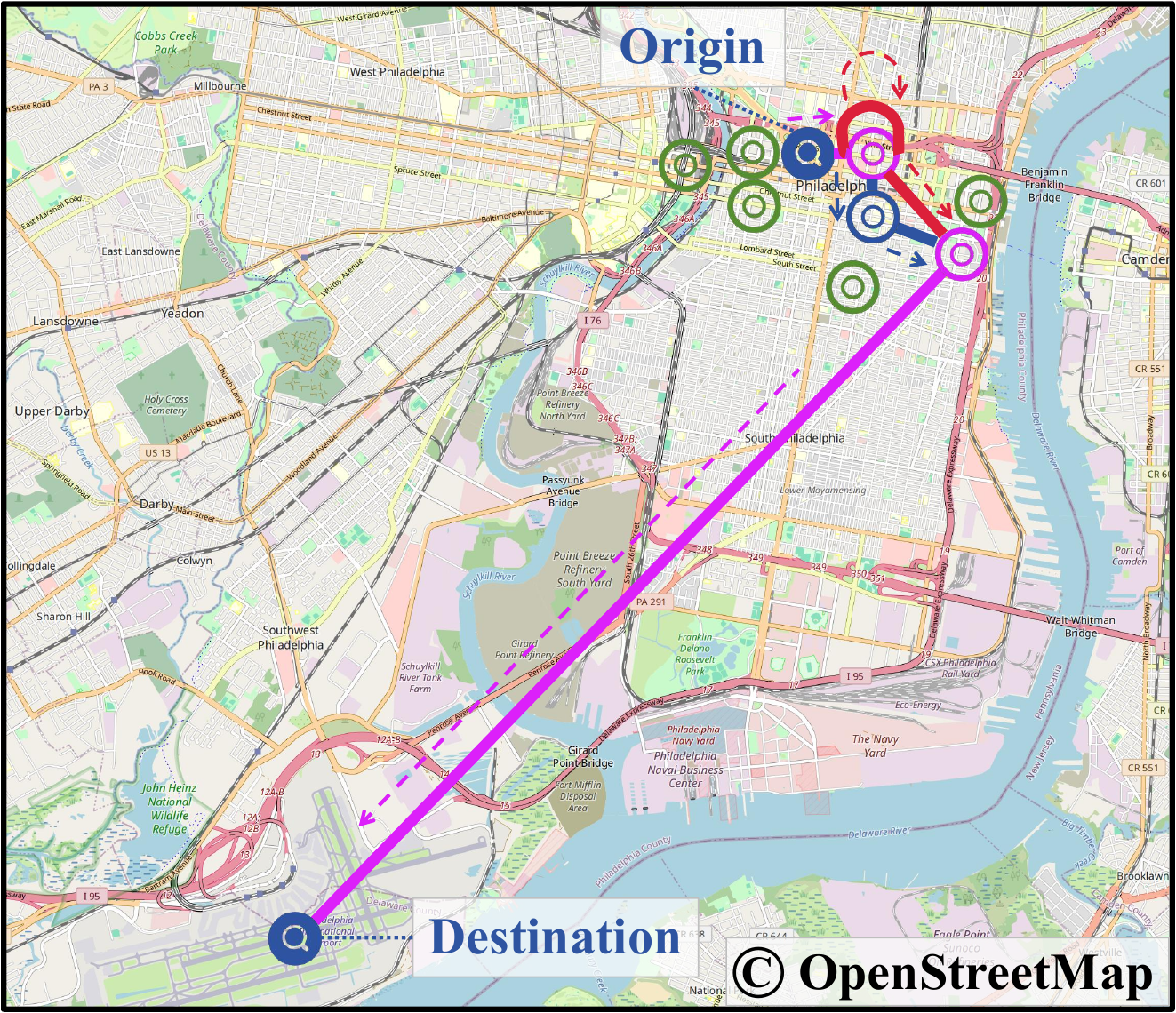}
		\caption{SPOT-Trip}
	\end{subfigure}
	\begin{subfigure}{0.245\linewidth}
		\centering
		\includegraphics[width=1\linewidth]{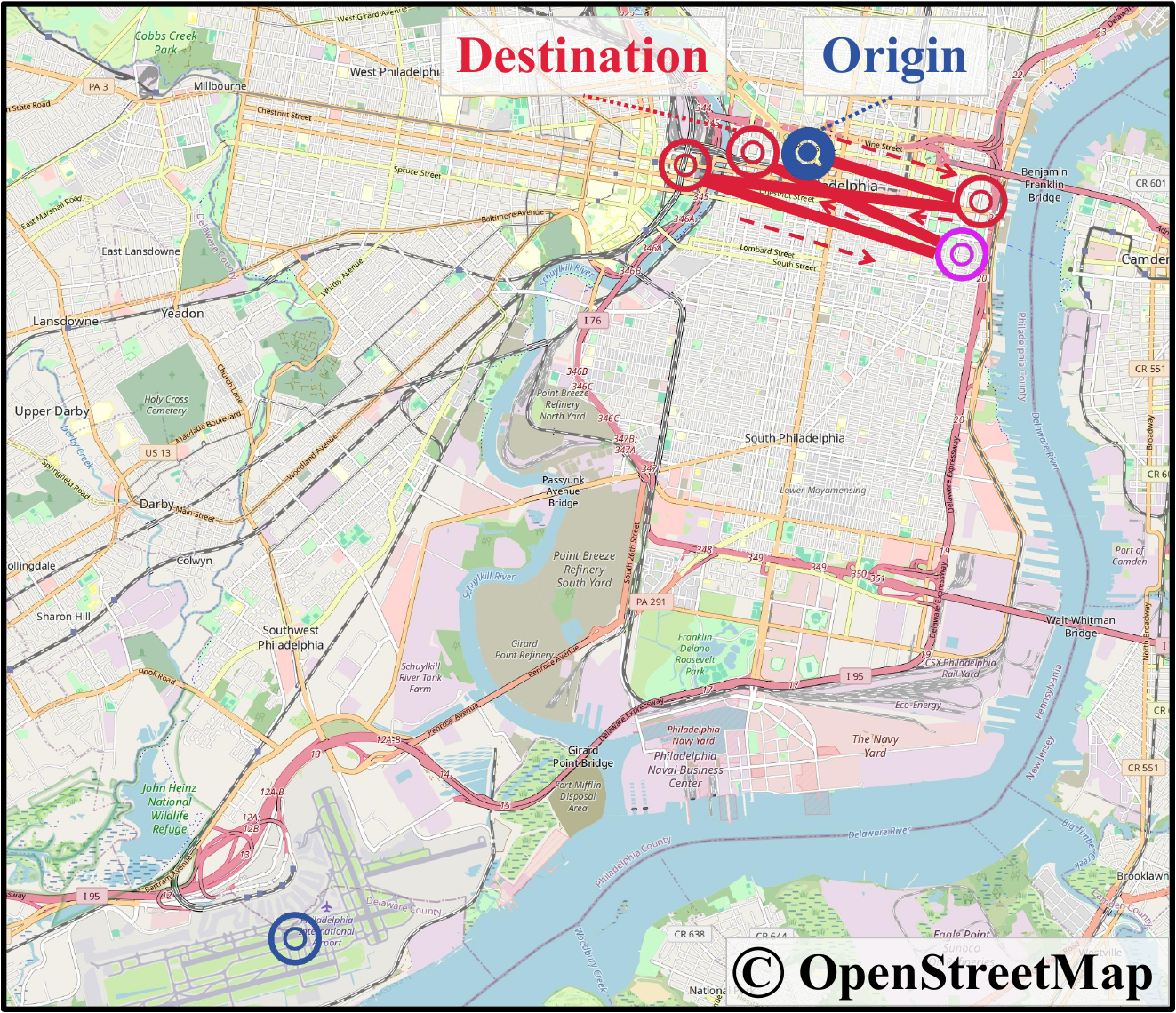}
		\caption{SPOT-Trip (O)}
	\end{subfigure}
	\begin{subfigure}{0.245\linewidth}
		\centering
		\includegraphics[width=1\linewidth]{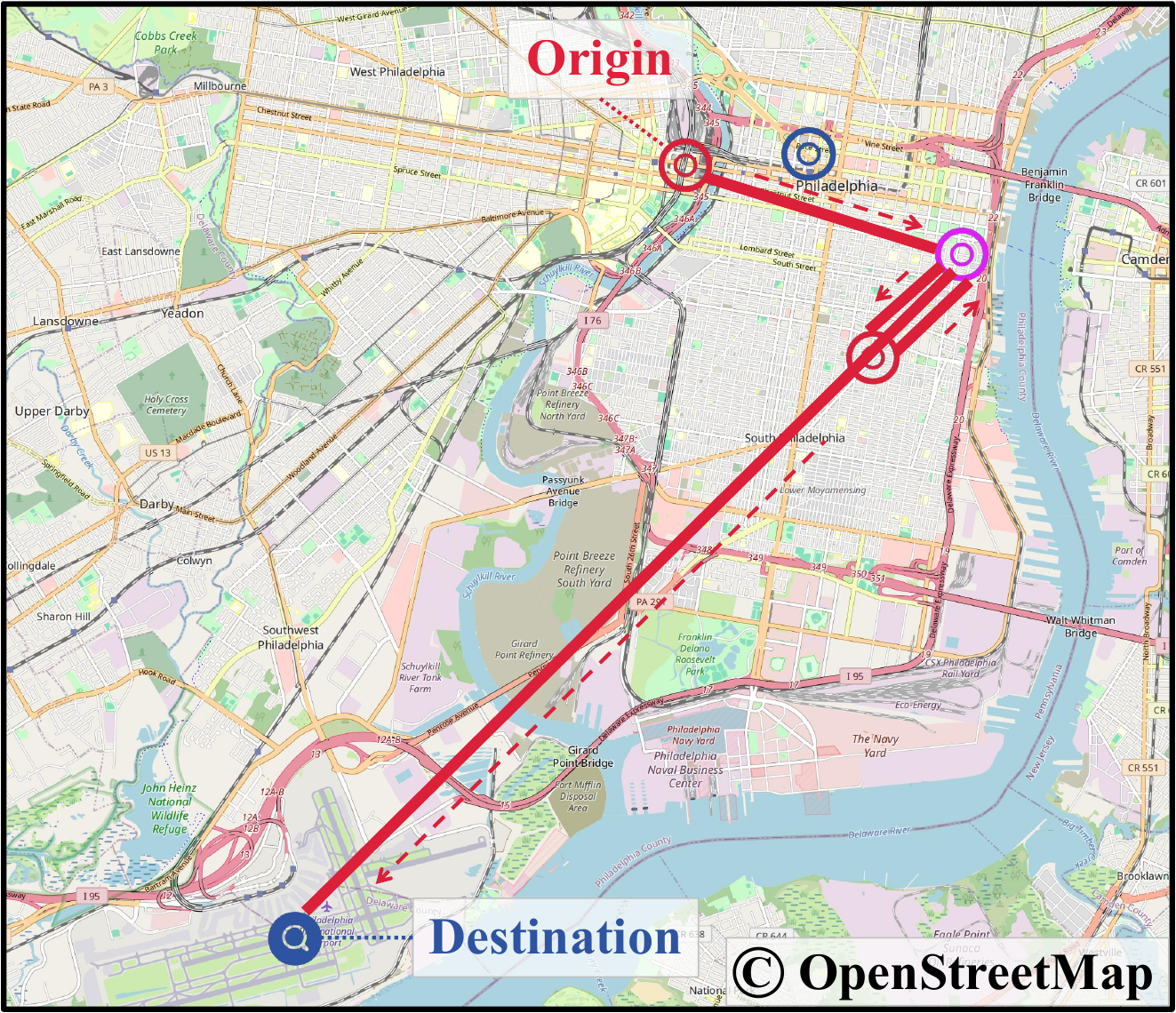}
		\caption{SPOT-Trip (D)}
	\end{subfigure}
	
	\vspace{0.3em}
	\textbf{Recommendation results for user 1090 on Foursquare}
	
	\vspace{0.8em}

	\begin{subfigure}{0.245\linewidth}
		\centering
		\includegraphics[width=1\linewidth]{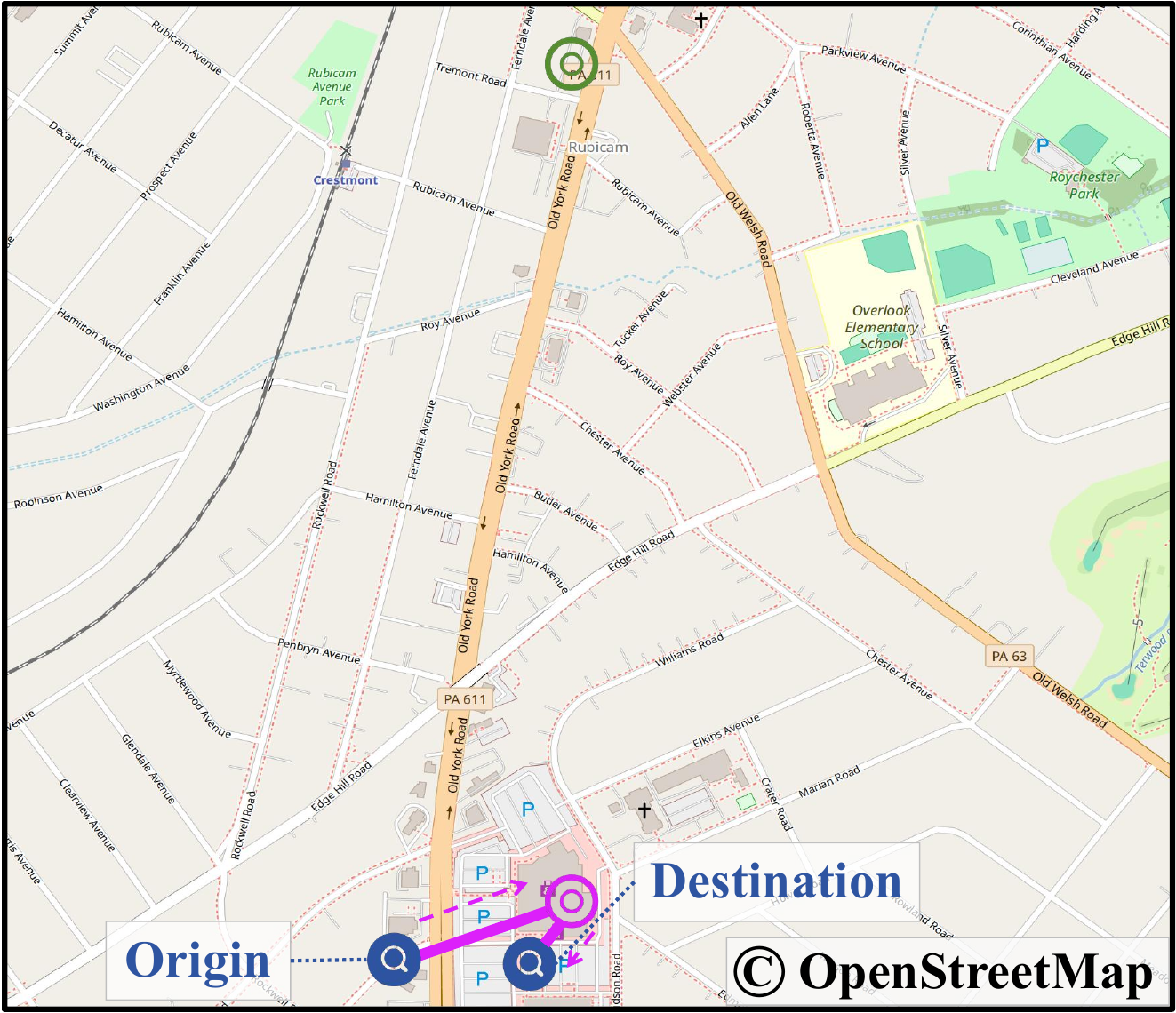}
		\caption{AR-Trip}
	\end{subfigure}
	\begin{subfigure}{0.245\linewidth}
		\centering
		\includegraphics[width=1\linewidth]{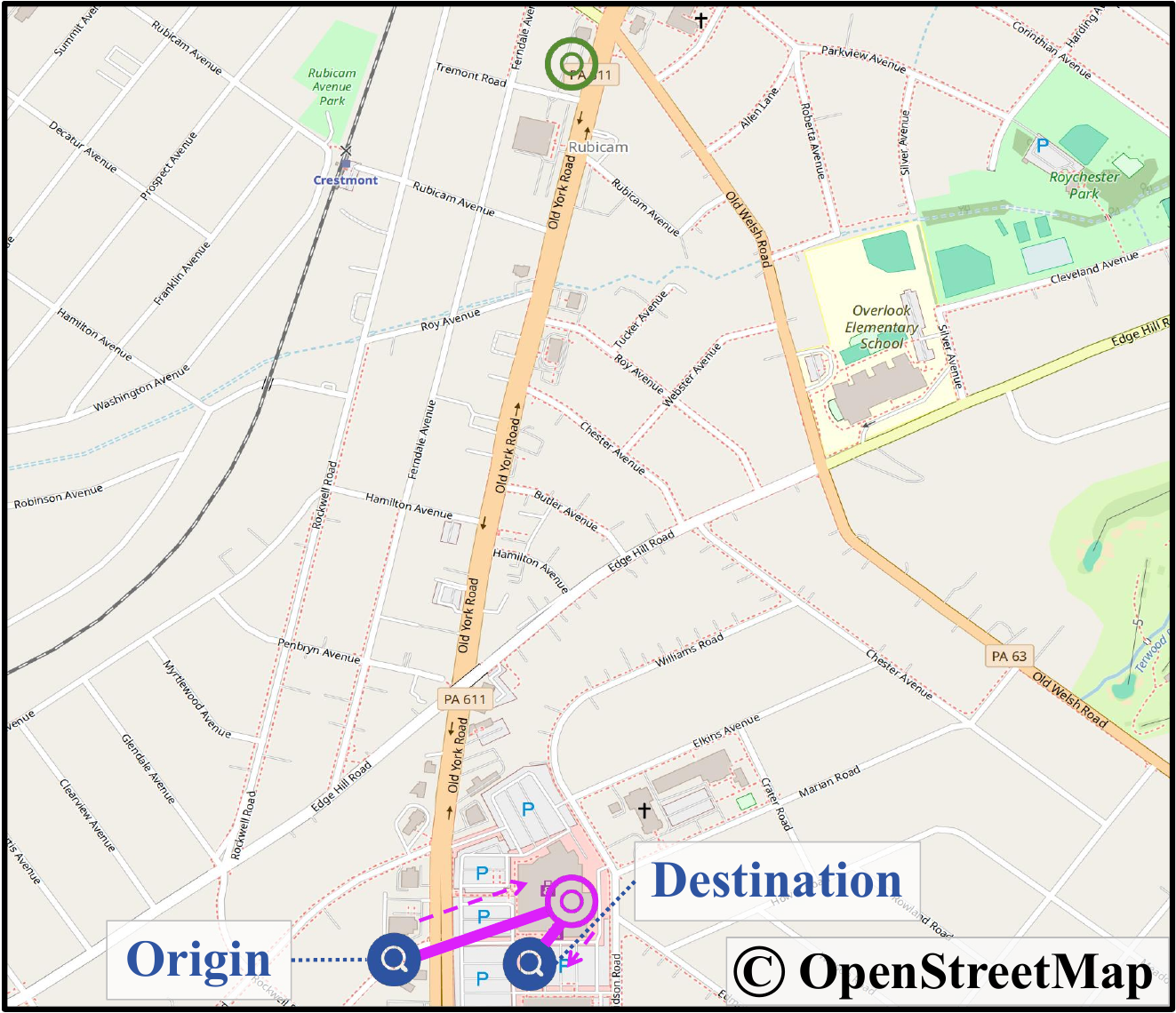}
		\caption{SPOT-Trip}
	\end{subfigure}
	\begin{subfigure}{0.245\linewidth}
		\centering
		\includegraphics[width=1\linewidth]{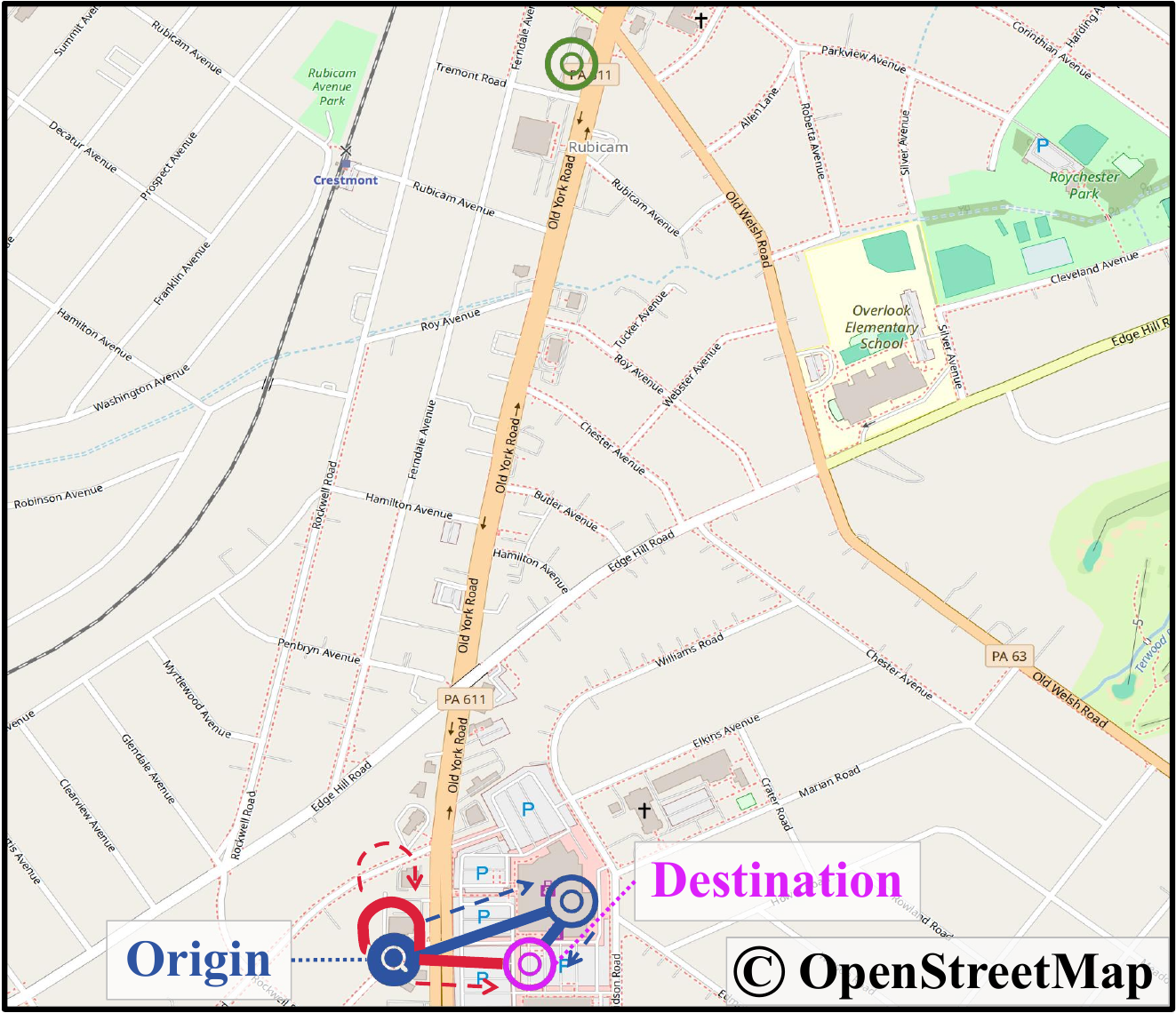}
		\caption{SPOT-Trip (O)}
	\end{subfigure}
	\begin{subfigure}{0.245\linewidth}
		\centering
		\includegraphics[width=1\linewidth]{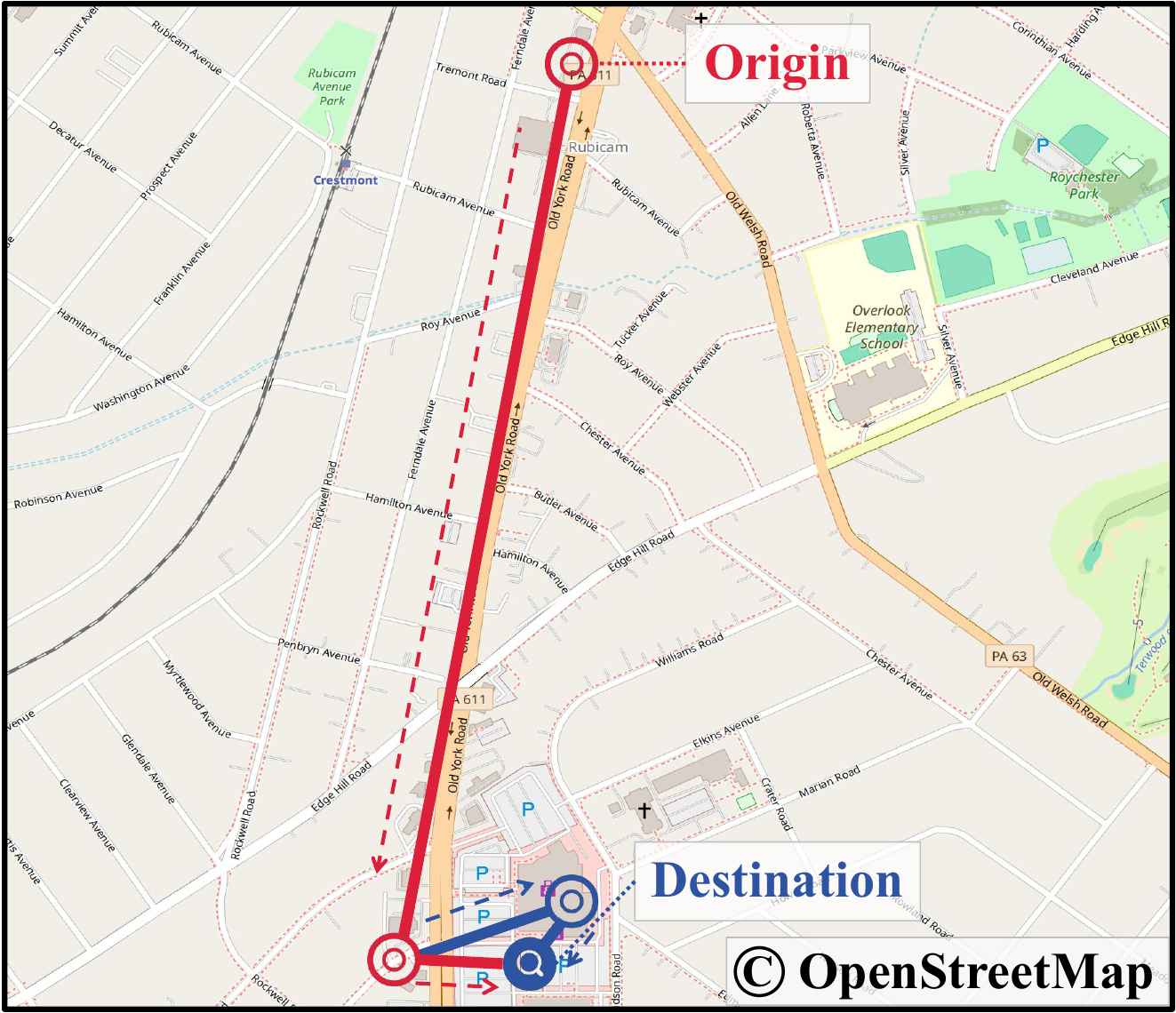}
		\caption{SPOT-Trip (D)}
	\end{subfigure}
	
	\vspace{0.3em}
	\textbf{Recommendation results for user 1544 on Yelp}
	
	\caption{Visualizations of recommendation results for users on Foursquare and Yelp.}
	\label{more case fig}
\end{figure}

We present more case studies for user 1090 on the Foursquare dataset and user 1544 on the Yelp dataset in Fig.~\ref{more case fig}. To avoid visual overlap and improve readability, some marks and lines have been reduced in size or omitted. The results display that as the number of intermediate query points increases (i.e., the user 1090's), AR-Trip’s recommendations become completely off-target, whereas SPOT-Trip tracks the ground truth closely, demonstrating our framework’s robustness to varying user query lengths. Notably, SPOT-Trip’s performance drops when handling queries that specify only an origin or only a destination. In future work, we will strive to enhance the method’s capability to meet such specific user requirements.

\section{Limitation} \label{limitations}
Despite the superior overall performance of SPOT-Trip, Fig.~\ref{more case fig} (b) and (g) exhibit repeated recommendations at proximate positions, which may undermine trip diversity. While AR-Trip~\cite{shu2024analyzing} incorporates a prior position matrix to partially mitigate this repetition, it is prone to error propagation: an incorrect POI recommendation at a position can mislead predictions at other positions, resulting in compounded inaccuracies. This cascading effect may ultimately degrade the quality of the recommended trip. In future work, we aim to develop more robust methods that simultaneously maintain recommendation accuracy and reduce redundant recommendations.

\end{document}